\documentclass{pasj00}

\begin{document}
\SetRunningHead{Deguchi et al.}{SiO Maser Survey off the Galactic Plane}
\Received{2009/09/28}
\Accepted{2009/02/27; PASJ Ver. 2.2;  Mar, 03, 2010}

\title{SiO Maser Survey off the Galactic Plane: A Signature of Streaming Motion}

\author{Shuji \textsc{Deguchi}\altaffilmark{1,2}, Tomomi \textsc{Shimoikura}\altaffilmark{1,3},}
\and
\author{Kazutaka \textsc{Koike}\altaffilmark{1,2}}

\altaffiltext{1}{Nobeyama Radio Observatory, National Astronomical Observatory,\\
              Minamimaki, Minamisaku, Nagano 384-1305}    
\altaffiltext{2}{Graduate University for Advanced Studies, 
National Astronomical Observatory, \\
Minamimaki, Minamisaku, Nagano 384-1305}
\altaffiltext{3}{current address: 
Department of Astronomy and Earth Sciences, Tokyo Gakugei University,\\
 Koganei, Tokyo 184-8501 \vspace{3mm}}

\author{ (PASJ 62 No.3, 2010 June 25 isuue in press)}

\maketitle

\begin{abstract}
A group of Mira variables in the solar neighborhood show unusual spatial motion 
in the Galaxy. To study this motion in much larger scale in the Galaxy, we newly surveyed   
134 evolved stars off the Galactic plane by SiO maser lines,
obtaining accurate radial velocities of 84 detected stars.
Together with the past data of SiO maser sources, we analyzed the radial velocity data of
a large sample of sources distributing in a distance range of about 0.3 -- 6 kpc 
in the first Galactic quadrant. 
At the Galactic longitudes between 20 and $40^{\circ}$, we found a group
of stars with large negative radial velocities, which deviate by more than 100 km s$^{-1}$ from
the Galactic rotation. We show that these deviant motions of maser stars
are created by periodic gravitational perturbation of the Bulge bar, and that the effect
appears most strongly at radii between corotation and outer Lindblad resonances.
The resonance effect can explain the displacement of positions from the Galactic plane as well. 
\end{abstract}


\section{Introduction} 
Stellar OH and SiO maser sources are powerful probes of Galactic structure and stellar evolution
\citep{hab06,deg08}. 
Radial velocity databases of these sources are useful to investigate dynamical motions 
of stars in the Galactic disk and Bulge \citep{izu95,sev01,deg04b,fuj06}. 
Because of a recent progress of studies of tidal streams surrounding our Galaxy
and in the solar neighborhood \citep{bel07,gri09}, one of urgent issues in this field is 
to separate a Bulge-bar resonance stream from tidal streams of relic dwarf galaxies 
in the radial velocity data.  
 
It has been known that the Galactic disk has two components, thick and thin;
the former has a thickness of 1 -- 2 kpc involving metal-poor stars and  kinematically peculiar stars,
while the latter has a thickness of about 300 pc involving young new populations.
A hypothesis that the thick disk is a relic of past merging processes 
has been proposed for the origin of thick disk \citep{hel99,nav04,hel06}. 

Moving groups in the solar neighborhood have also been known, and
they are considered to be fossils keeping dynamical information of their birth. 
Two famous examples are the Arcturus and Hercules groups of stars \citep{egg96}; 
the former is a group of metal poor stars with a coherent spatial motion 
with a $\sim 100$ km s$^{-1}$ lag to the Galactic rotation, 
and the latter is a stellar group with a heterogeneous mixture of metal abundances and with a smaller rotational lag. 
Spatial motions of these moving groups are well investigated optically 
based on the Hipparcos (proper motions) and
the RAVE (the radial velocities) databases. However, these investigations have a limitation
of distance up to about 1 kpc due to lack of proper motion data. Because of heterogenous metal abundances of member stars,
the origin of the Hercules stream is attributed to a resonance of the bar-like Bulge
\citep{ben07}. \citet{fea00} investigated an outward motion of short-period Mira variables near the Sun,
and attributed it to the resonance effect of the Bulge bar. 

In this paper, we reinvestigate the radial velocity data of SiO maser sources
toward the region of $l=20$ -- 60$^{\circ}$, and $-30<b<60^{\circ}$ (excluding the 
galactic plane, $|b|<3^{\circ}$). In this region,
\citet{deg07} found a group of SiO maser stars with large negative velocities
($v_{\rm LSR} \sim -70$ km s$^{-1}$), which may be attributed to a resonance effect of the Bulge bar or a relic streaming motion.
However, the previous off-plane SiO search was 
aimed to find distant debris stars such as those in the Sgr dwarf streams, the search 
was somewhat shallow in depth except toward Sgr dwarf streams. 
Therefore, we have made a new sensitive observation by SiO maser lines toward the thick disk, 
and have added more data in radial velocity database. 
Here, we analyze kinematics of this SiO maser star stream, and test if 
this stream is originated by the gravitational perturbation of the Bulge bar.   
 


\section{Observational results}
The observations were made with the 45m radio telescope at Nobeyama in 2009 April and May
by 
the SiO $J=1$--0 $v=1$ and 2 transitions at 43.122 and 42.821 GHz, respectively.
Cooled HEMT receiver (H40)  was used for the 43 GHz observations
with acousto-opt spectrometer arrays with the 40 and 250 MHz bandwidths (velocity resolution
of about 0.3 and 1.8 km s$^{-1}$, respectively).
The overall system temperature was about 180 --- 250 K for the SiO observations depending on weather conditions.
The half-power beam width (HPBW) of the telescope was about 40$''$ at 43 GHz. 
The conversion factor of the antenna temperature to the flux density was about 2.9 Jy K$^{-1}$. 
All of the observations were made by the position-switching mode. 
Further details of observations using the NRO 45-m telescope have been
described elsewhere \citep{deg00}.

The sample for SiO searches was chosen 
in the area within $20^{\circ}<l<60^{\circ}$ and $|b|<45^{\circ}$ (excluding $|b|<3^{\circ}$)
by the selection criteria which have been established well in the past SiO surveys.
The mid-infrared objects brighter than 3 Jy at 12 $\mu$m, 
and the color $-0.5 <C_{12} [\equiv log(F_{25}/F_{12})] \lesssim 0.2$ were selected from IRAS point source catalog
\citep{joi88},
where  $F_{12}$ and $F_{25}$ are the IRAS flux densities in the 12 and 25 $\mu$m bands, respectively. 
The MSX bands C and E \citep{ega03} were also consulted for the $|b| \lesssim 6^{\circ}$ sources.
Then, we checked whether or not the MIR objects
have a NIR counterpart in the 2MASS catalog \citep{cut03} with a customary selection
criteria in our SiO maser searches \citep{deg04b}: $K<9$, and $H-K>0.9$ for an initial sample.
All the  objects in the present sample have 2MASS counterpart brighter than $K=8.2$ mag. 
These sources are supposedly late-type (AGB or post-AGB) stars with circumstellar dust 
in a color-temperature range between 250 and 1000 K. Excluding previously observed objects, we finally selected 
 about 150 candidates which satisfied above criteria in this sky area. However, because of time restriction of observations,
 we completed half of these sources, for which we consumed all of the objects above $F_{12}=5$ Jy.
Furthermore, we added bright objects for backup (for bad weather condition), which involves slightly 
bluer sources in $H-K$ but not surveyed before. We added these additional objects to our results 
for completeness. 

Observational results are summarized in tables 1 and 2 for SiO detection and no detection, respectively.
The observed spectra of the SiO $J=1$--0 $v=1$ and 2 transitions are shown in figure 1a -- 1e for the detections.
Table 3 summarizes infrared properties of the observed sources.
For a distance measure, we use the corrected K magnitude for interstellar and circumstellar reddening,
\begin{equation}
K_c= K - A_K/E(H-K)\times [(H-K)-(H-K)_0],
\end{equation}
where we use $A_K/E(H-K)=1.44$ and $(H-K)_0 =0.5$, which is appropriate for M5III stars
\citep{fuj06}. The corrected $K_c$ is listed at the 6th column in table 3.
A typical Mira star with a period of about 450 d located at the Galactic center (distance of 8 kpc)
without extinction has $K_c=6.43$  \citep{gla95}.
We will use this value to estimate distances in the next section.
Because SiO maser stars are mostly miras,  the period--$K_c$ relation [at 8 kpc; \citet{gla95}] gives a relatively smaller dispersion 
of  about 1 magnitude in average $K_c$ for the Bulge SiO maser stars [with an average period of $\sim 490$d $\pm$ 130d ;  
see figure 3 of \citet{deg04b}]. Because the  single-epoch 2MASS photometric magnitude may differ from the average value by about 1 magnitude 
 [e.g., Figure 11 of \citet{mes04}], and furthermore absolute K magnitude depends on the spectral type of a star  \citep{wai92},
we deduce that  the error in distance in the present paper is a factor of more than 2. 

Figure 2 shows the $K$ -- $H-K$ and log($F_{12}$) -- $C_{12}$ diagrams for the observed sources.
These panels show that the color-selection criteria described above can extract the SiO emitting objects quite effectively
from the infrared star catalogs.
Figure 3 shows histograms of log($F_{12}$) and $K_{c}$ for the SiO detection and no detection. 
The SiO detection rate are quite high ($\sim 80$ \%) for bright infrared objects in $F_{12}$ and $K_{c}$ terms,
but it decreases with decreasing infrared fluxes. Beyond $K_{c}>5.5$, no detection surpasses
the detection because of the large distance. These diagrams show properties similar to those 
made in the previous surveys in the Galactic plane (for example, \cite{deg04b}), 
and assure that the present survey off the Galactic plane was made appropriately.

\begin{figure*}
  \begin{center}
    \FigureFile(150mm,180mm){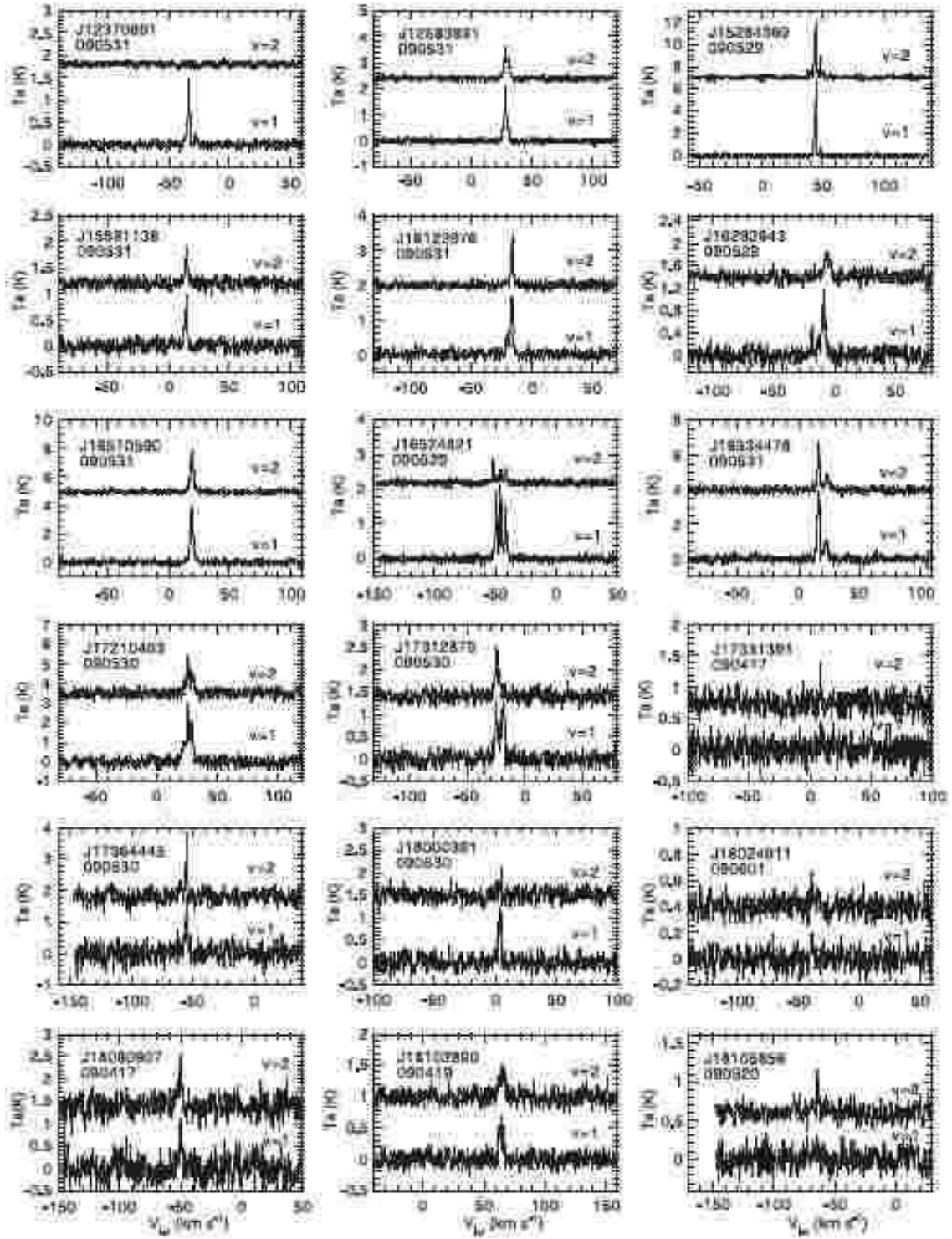}
  \end{center}
  \caption{a. SiO $J=1$--0 $v=1$ and 2 spectra. The source name (Jhhmmssss format) and the observed date (yymmdd format) are shown
at the upper left of each panel.  
}\label{fig: spectra-1a}
\end{figure*}
\setcounter{figure}{0}
\begin{figure*}
  \begin{center}
    \FigureFile(150mm,180mm){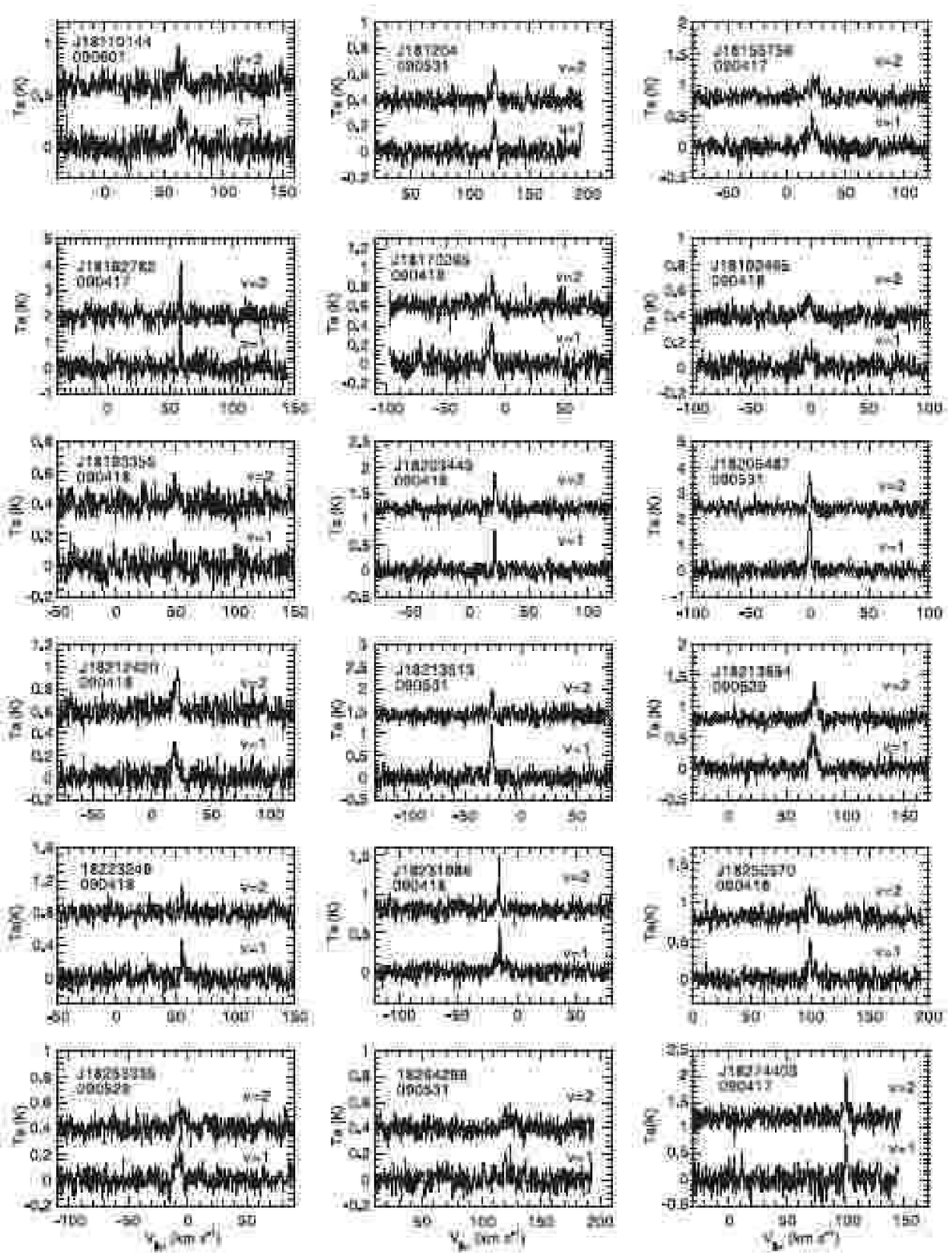}
  \end{center}
  \caption{b. SiO $J=1$--0 $v=1$ and 2 spectra. The source name (Jhhmmssss format) and the observed date (yymmdd format) are shown
at the upper left of each panel.  
}\label{fig: spectra-1b}
\end{figure*}
\setcounter{figure}{0}
\begin{figure*}
  \begin{center}
    \FigureFile(150mm,180mm){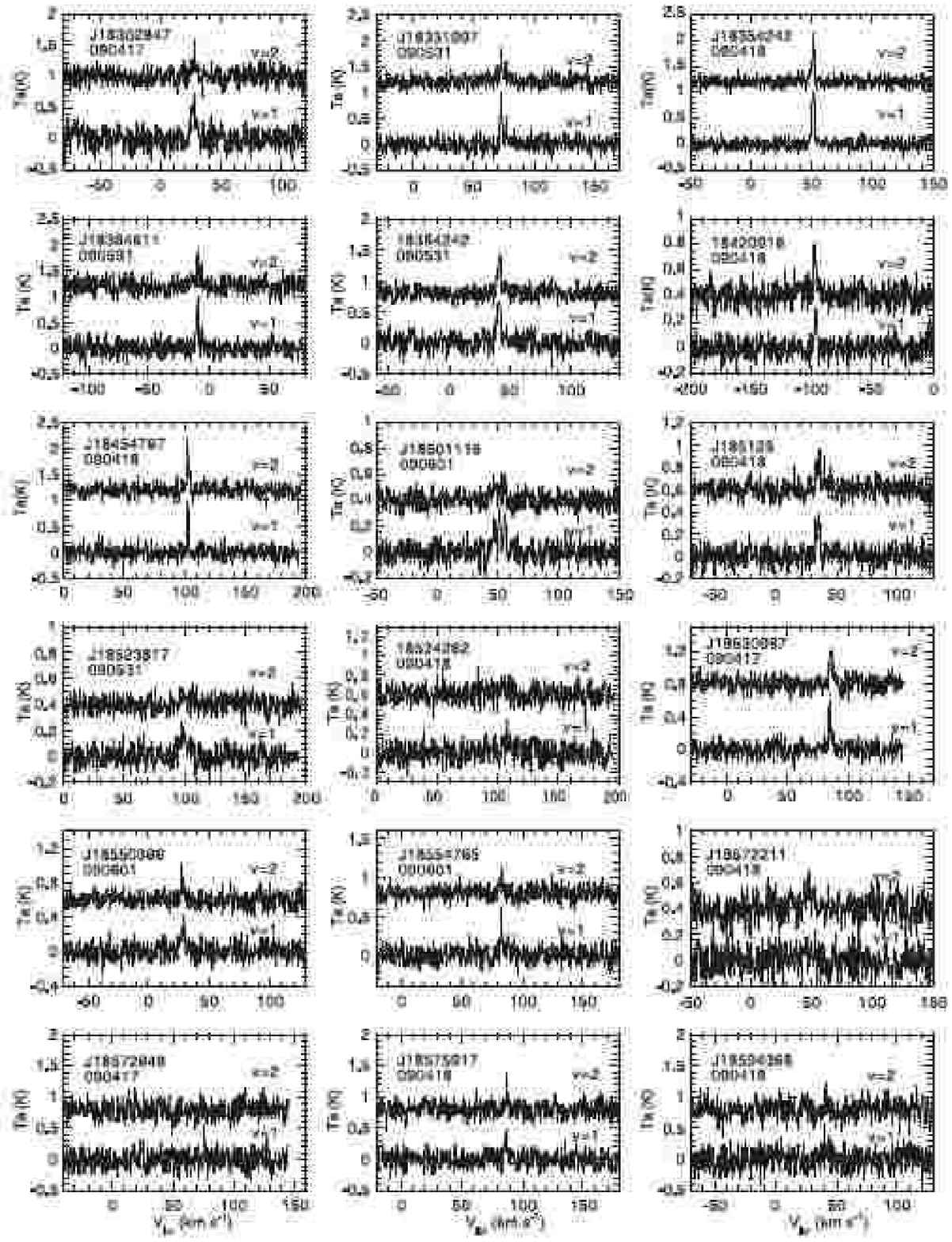}
  \end{center}
  \caption{c. SiO $J=1$--0 $v=1$ and 2 spectra. The source name (Jhhmmssss format) and the observed date (yymmdd format) are shown
at the upper left of each panel.  
}\label{fig: spectra-1c}
\end{figure*}
\setcounter{figure}{0}
\begin{figure*}
  \begin{center}
    \FigureFile(150mm,180mm){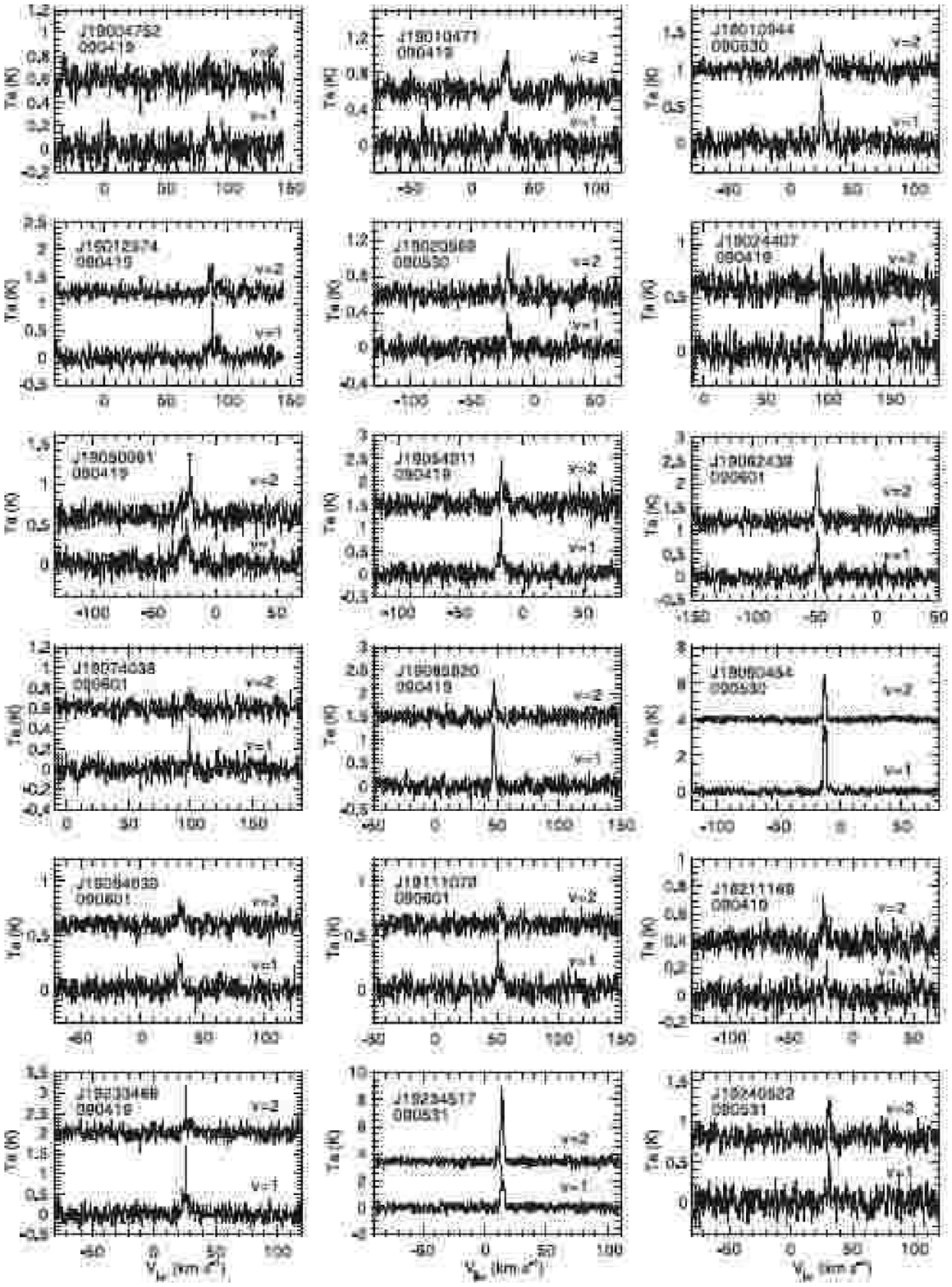}
  \end{center}
  \caption{d. SiO $J=1$--0 $v=1$ and 2 spectra. The source name (Jhhmmssss format) and the observed date (yymmdd format) are shown
at the upper left of each panel.  
}\label{fig: spectra-1d}
\end{figure*}
\setcounter{figure}{0}
\begin{figure*}
  \begin{center}
    \FigureFile(150mm,180mm){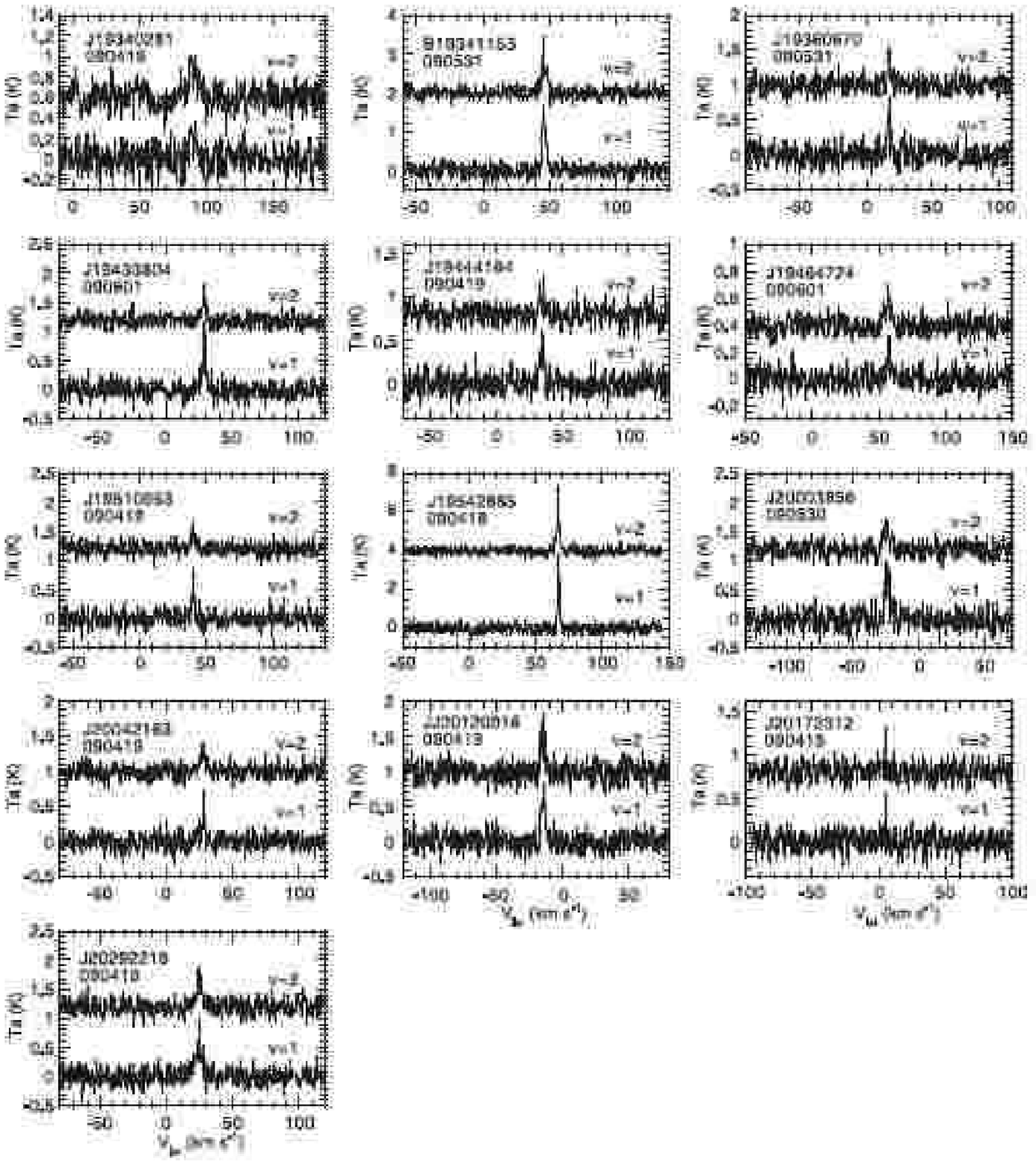}
  \end{center}
  \caption{e. SiO $J=1$--0 $v=1$ and 2 spectra. The source name (Jhhmmssss format) and the observed date (yymmdd format) are shown
at the upper left of each panel.  
}\label{fig: spectra-1e}
\end{figure*}
%
\setcounter{figure}{1}
\begin{figure}
  \begin{center}
    \FigureFile(80mm,60mm){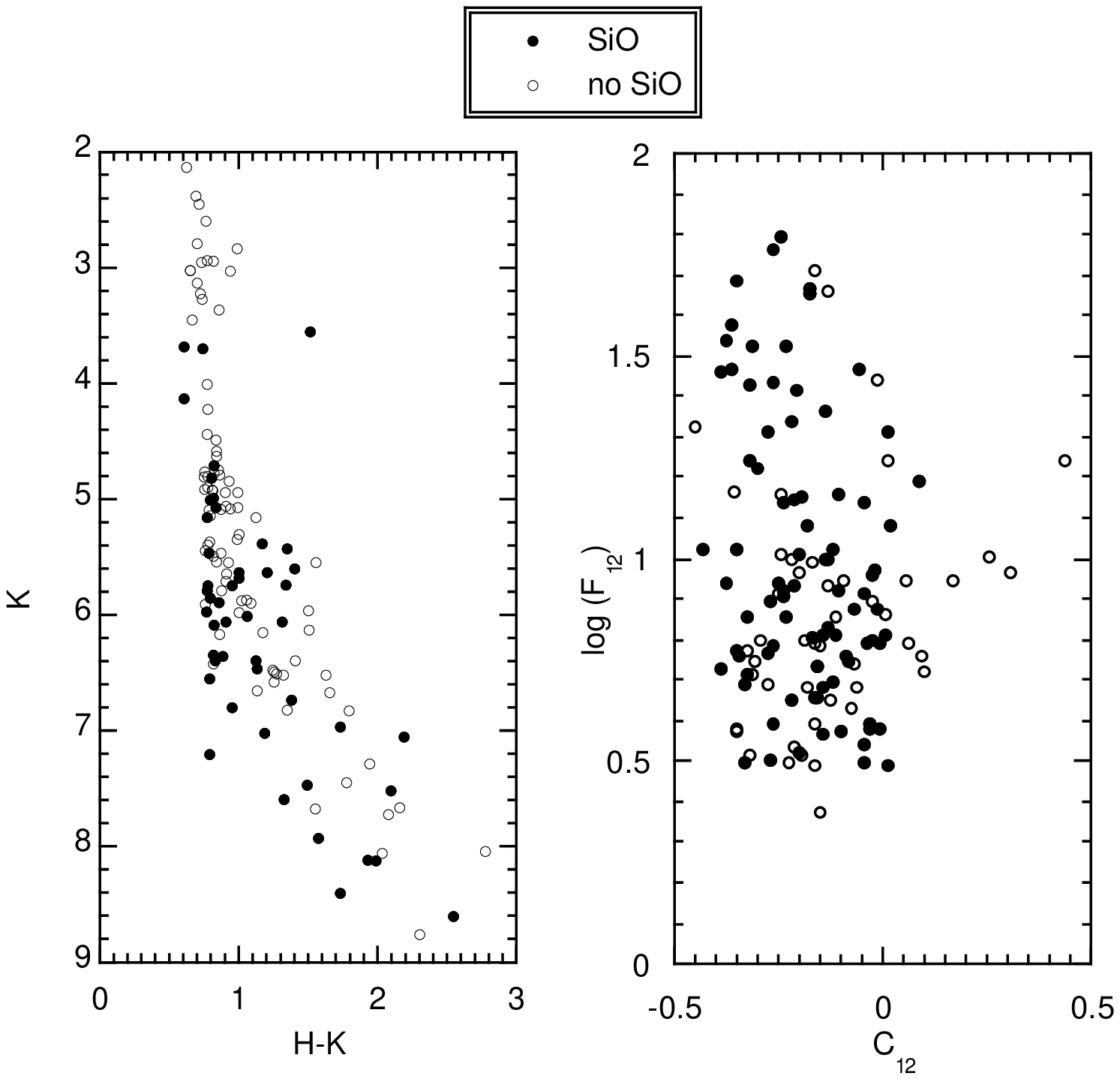}
  \end{center}
\caption{NIR and MIR  color-magnitude diagrams.
}\label{fig: magnitude-color}
\end{figure}


\begin{figure}
  \begin{center}
    \FigureFile(80mm,60mm){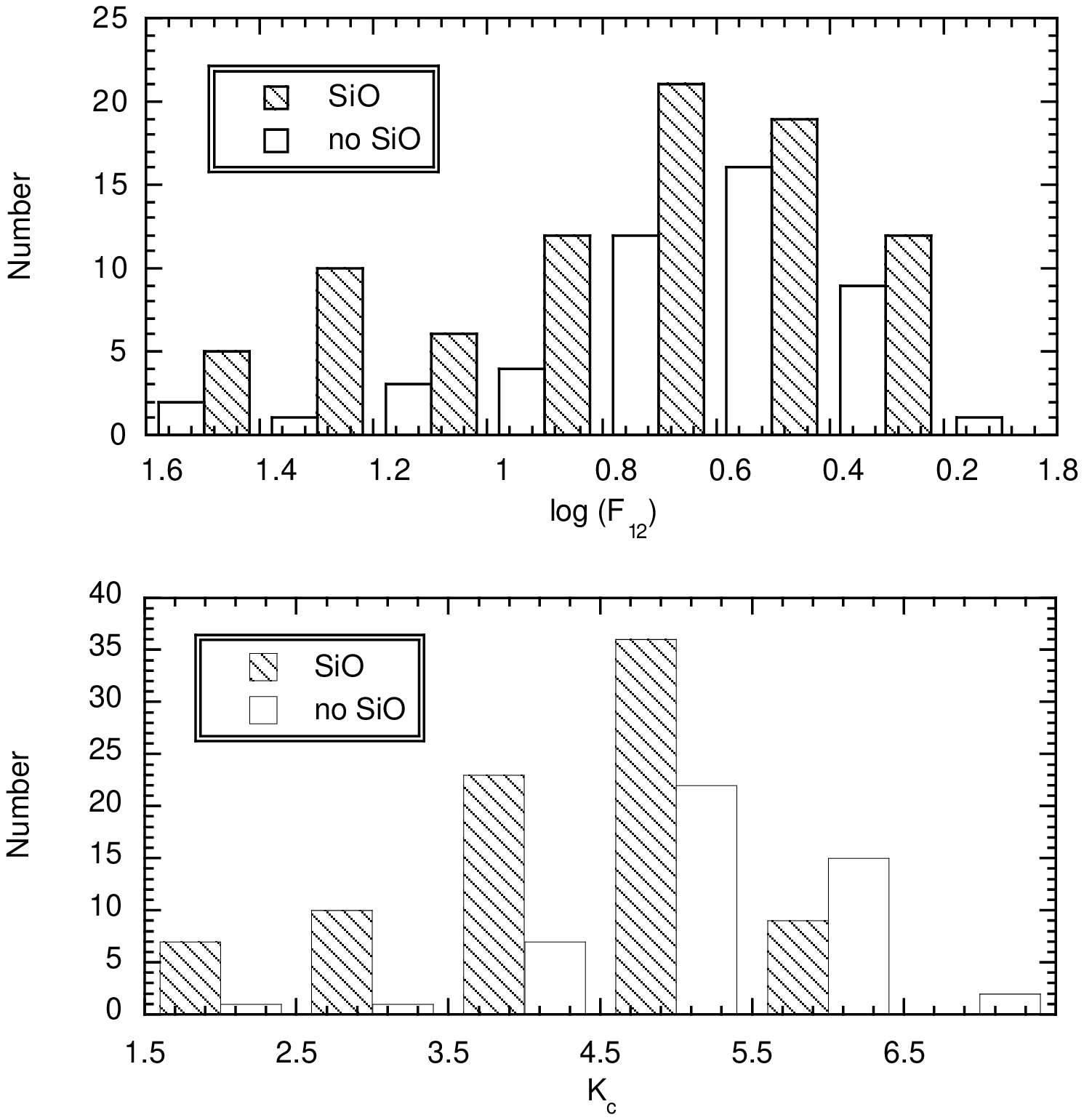}
  \end{center}
  \caption{histograms of log($F_{12}$) and $K_{c}$ for SiO  detections and no detections.
}\label{fig: N-mag}
\end{figure}


\section{Discussion}

\subsection{Selection of the candidates in streaming}
Figure 4 shows a longitude-velocity diagram for the SiO  sources found in this work (filled circles)
and those in our past SiO surveys (unfilled circles).    
We can see a considerable increase of SiO detections in this sky region.
 A large vertical (velocity) spread of sources below $l=17^{\circ}$ in figure 4
is attributed to the Bulge stars.
At $l\sim 17^{\circ}$, there is a gap in the spread of radial velocities.   
Beyond $l=18^{\circ}$, the velocity spread increases again.  The lack of stars beyond $V_{\rm LSR} \sim 50$ km s$^{-1}$ at the gap
indicates that it is the edge of the Galactic Bulge (at least, for the peanut-shape thick Bulge with $|b|>3^{\circ}$).
Therefore, in this paper, we specially pay attention to the sources with $v_{\rm LSR}<0$ in the region of $l=18$--$40^{\circ}$.
Because the Galactic circular rotation gives positive $v_{\rm LSR}$ in the range $l=0$--$90^{\circ}$ (in the solar neighborhood),
it is hard to separate any streaming motions, if present, at the $v_{\rm LSR}>0$ side in this diagram. 
Note that the stars on the solar circle
fall on the $v_{\rm LSR}=0$ line in the $l$-$v$ diagram (if they circularly rotate around the Galactic center with 
the circular velocity same as that of the Local Standard of Rest).  
Two concentrations of the $v_{\rm LSR}<0$ stars are seen in figure 4: one around $l=20$--$25^{\circ}$, and another $l=30$--$40^{\circ}$.
Beyond $l=45^{\circ}$, we also see a mild scatter of stars with large negative velocities.  However,
 at the range beyond $l=45^{\circ}$,  the stars in the distant spiral arms ($D>11$ kpc at $l=45^{\circ}$)   fall
 outside the solar circle and  have the negative $v_{\rm LSR}$ in the $l$-$v$ diagram. Therefore, it is more or less difficult to separate
the streaming candidates beyond $l=45^{\circ}$ unless distances are accurately known.  	
 
\begin{figure}
  \begin{center}
    \FigureFile(80mm,60mm){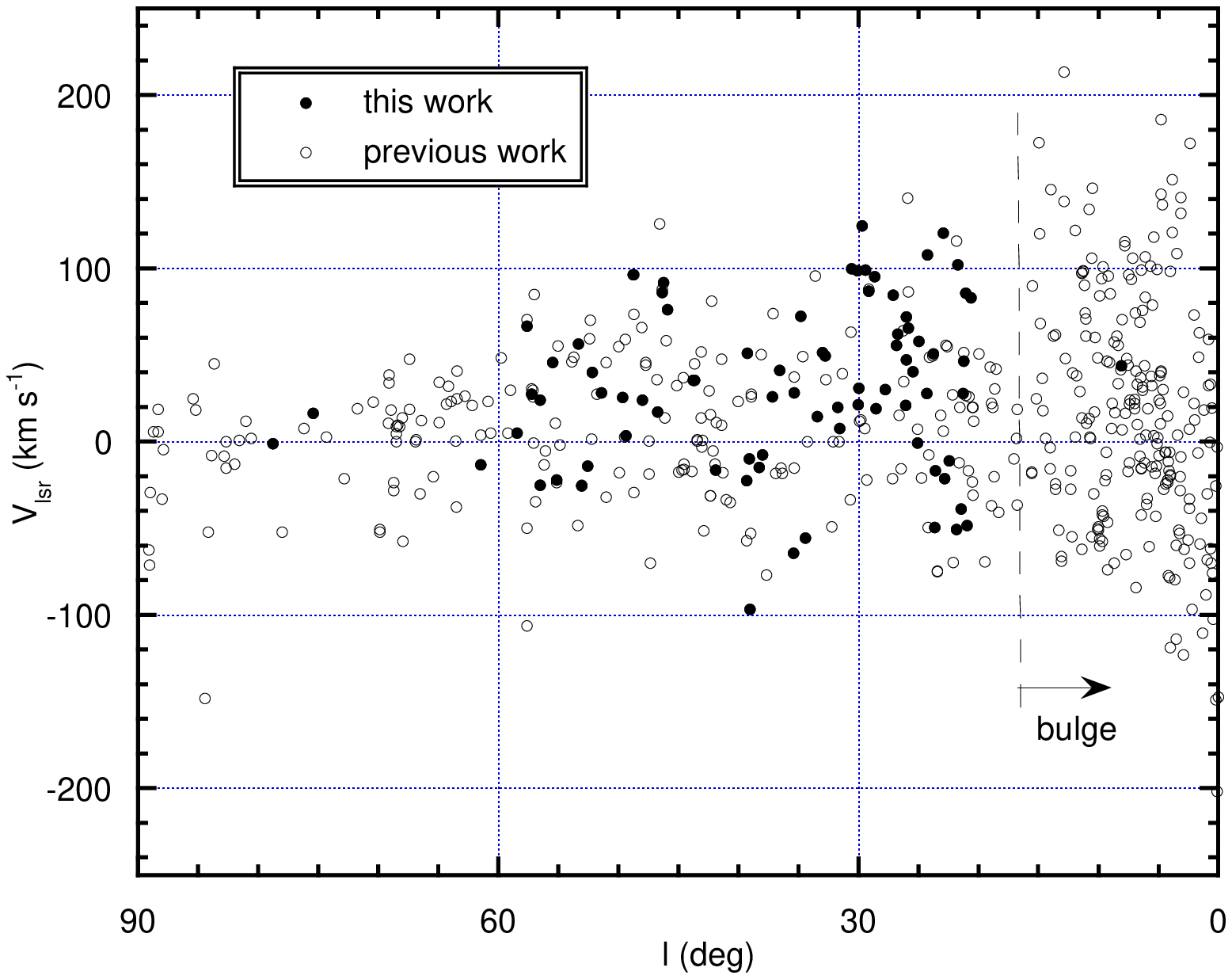}
  \end{center}
  \caption{Velocity-longitude diagram of SiO maser sources at $|b|>3^{\circ}$.  
Filled and unfilled circles indicate the SiO sources detected in this and the previous works, respectively.
 The vertical dashed line at $l=17^{\circ}$ indicates a gap of the velocity spread (see text).
}\label{fig: fig4}
\end{figure}

\begin{figure}
  \begin{center}
    \FigureFile(80mm,60mm){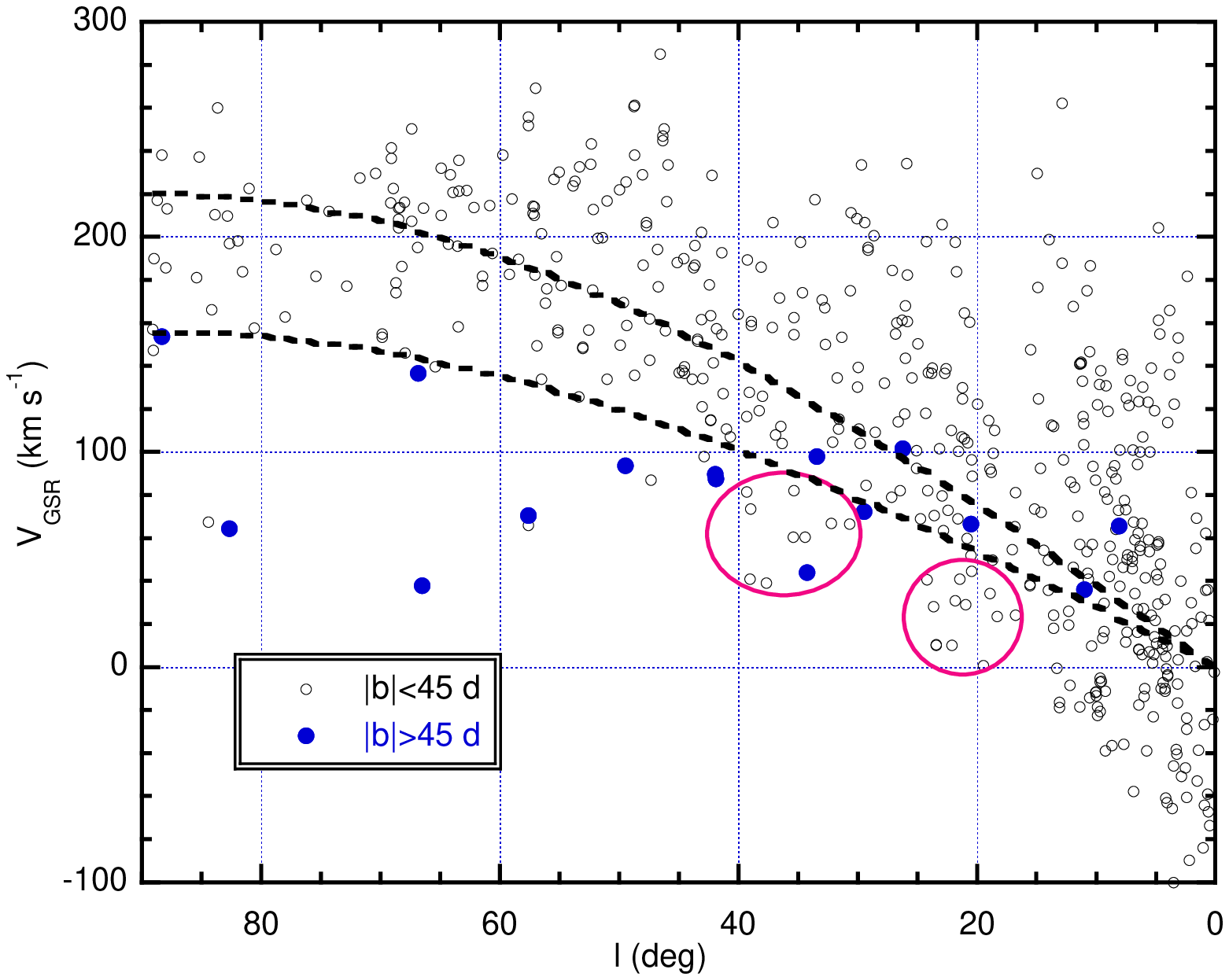}
  \end{center}
\caption{ Velocity-longitude diagram ($V_{\rm GSR}$ -- $l$). Filled and unfilled circles indicate
the objects at high ($|b|>45^{\circ}$) and low ($|b|<45^{\circ}$) galactic latitudes. 
 The broken curves indicates the $v_{\rm LSR}=0$ line (figure 4)
in the cases of $|b|=0^{\circ}$ (upper) and $45^{\circ}$ (lower). 
}\label{fig: fig5}
\end{figure}

\begin{figure}
  \begin{center}
    \FigureFile(60mm,100mm){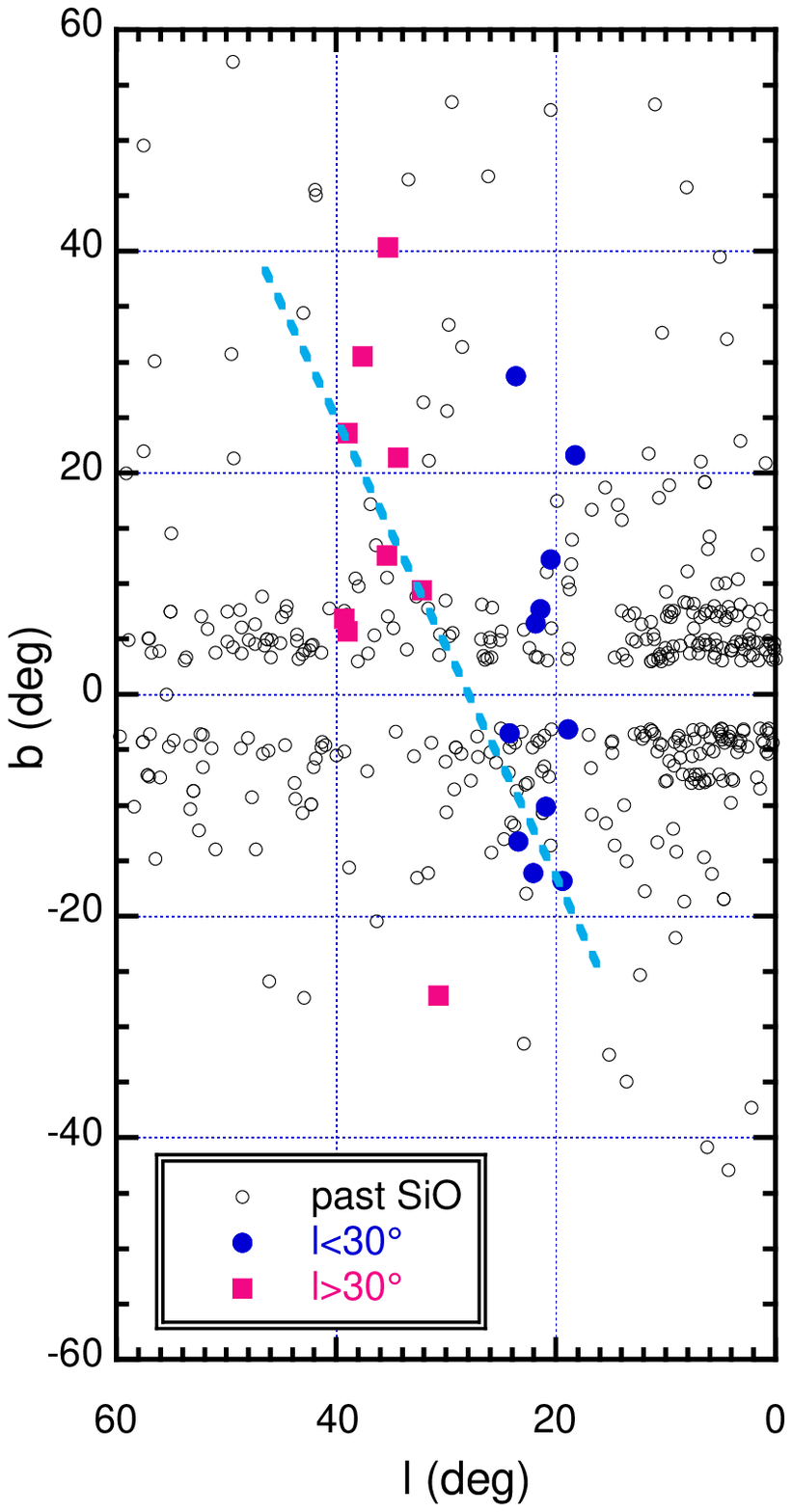}
  \end{center}
  \caption{Positions of the SiO sources in the Galactic coordinates.
  Filled and unfilled symbols indicate the deviant candidate and the usual disk star,
  respectively. The broken line indicates a possible orbital locus if they are interpreted as
  a debris stream \citep{deg07}. 
}\label{fig: fig6}
\end{figure}
%
\begin{figure}
  \begin{center}
    \FigureFile(80mm,50mm){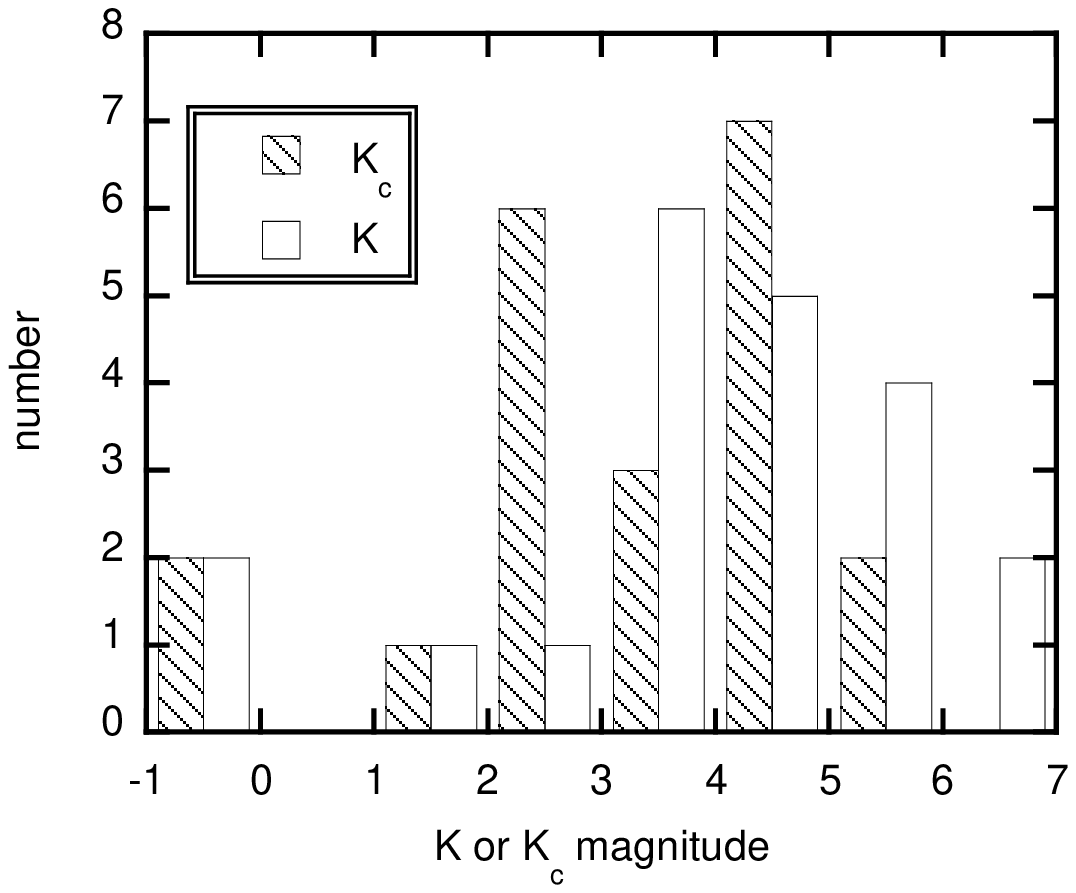}
  \end{center}
  \caption{
  Histogram of $K_c$ (shaded) and $K$ (unshaded). 
}\label{fig: fig7}
\end{figure}

 
For simplicity, we restrict the later discussion only to the stars with large negative velocities
($v_{\rm LSR} \lesssim -40$ km s$^{-1}$) in the longitude range between $l=18$ and $40^{\circ}$.
However, it is not clear how to select the streaming candidates from usual
stars with large random motion. 
To make the separation as definitive as possible, and to minimize the effects due to Galactic longitudes and latitudes, 
we made the plot (figure 5) by the velocity with respect to the Galactic Standard of Rest (GSR), where
\begin{equation}
v_{GSR}=v_{\rm LSR} + V_{\phi}\ {\rm sin(} l)\ {\rm cos(} b) .
\end{equation}        
Here, we use the LSR rotational velocity around the Galactic center,  $V_{\phi}= 220$ km s$^{-1}$.
We apply, for the later use, the standard solar motion of 20 km s$^{-1}$ in the direction of $R.A.=18^{h}00^{m}00^{s}$ and
$Dec. =30^{\circ}00'00''$ (1900.0)  ({\it radio definition}) with respect to the Local Standard of Rest
and the Sun -- Galactic-center distance of 8.0 kpc.
The two broken curves in figure 5 indicates the $v_{\rm LSR}=0$ line in figure 4 
in the cases of $|b|=0$ (upper curve) and $45^{\circ}$ (lower curve). 
For the current purpose of separating streaming candidates, we selected the stars under the lower broken curve
between $l=18$ and $40^{\circ}$ excluding the objects with $|b|>45^{\circ}$; the candidates are in two ellipses. 
Figure 6 shows the sky positions of the candidates (filled circle) 
and else (open circles) in the Galactic coordinates. 
Because a boundary between the stars in random motion and the stars in streaming is not clear at present, 
we have chosen all the likely candidates, which are in the ellipses in figure 5, and investigate the nature of this subsample.
In fact, we will see in the next section that all of these objects are highly likely objects in a stream. 

The distribution of these candidate stars in the sky is shown in figure 6. It spreads widely in latitude
($\sim 60 ^{\circ}$), but not much in longitude ($\sim 20^{\circ}$).	   
Infrared properties of these candidate stars are summarized in table 4.  We created
the magnitude-color and color-color diagrams of the candidates 
in near and middle-infrared bands (similar to figure 2), and compared the distribution with that of noncandidates. 
The candidates and noncandidates 
seem to have no clear difference in distribution in these diagrams,
suggesting that the physical properties of the streaming candidates 
are not very different from those of normal disk SiO sources.

\subsection{Kinematic property of the candidate stars and model fittings}
Figure 7 shows histograms of apparent and corrected K magnitudes ($K$ and $K_c$ ) for the candidates.
The histogram of $K_c$ shows triple peaks, indicating that the candidates
can be separated into three groups according to their magnitudes: the bright group
with $K_c=-1$ -- 0, the middle group with $K_c=1$ -- 4, and the faint group with  $K_c>4$.
Grouping into three may not be very meaningful due to statistical errors. 
However, let us separate the sample into three for convenience, and see if
any useful properties reveal.

\begin{figure}
  \begin{center}
    \FigureFile(70mm,130mm){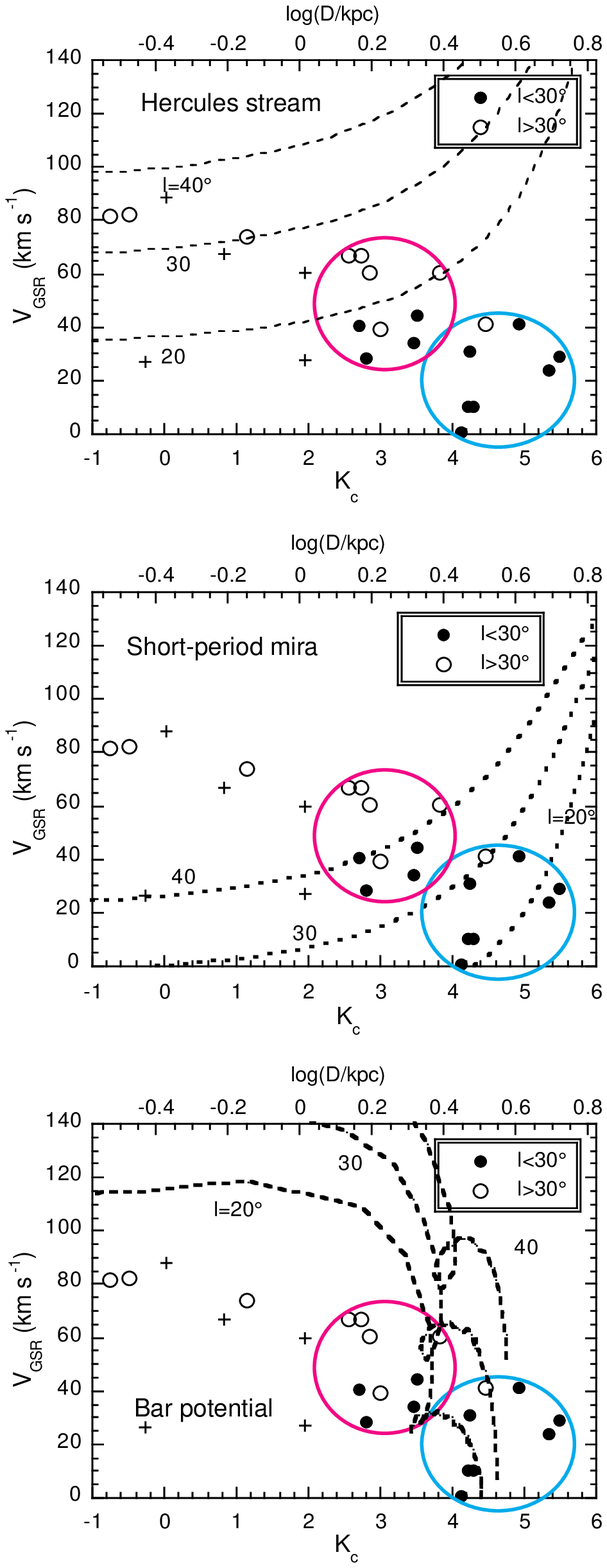}
  \end{center}
  \caption{Comparison of the velocity fittings for the diagram of $v_{\rm GSR}$ against $K_c$, 
  where corrected magnitude, $K_c$, can be converted to a distance, as shown in the upper axis. 
   Open and filled circles indicate the candidate SiO sources above and below $l=30^{\circ}$, 
  and crosses indicate short-period miras at $20<l<70^{\circ}$ [ listed as Group 2 in Table 2 of \citet{fea00}.
  Three broken curves in the upper panel show plots of GSR velocities at $l=20$, $30$ and $40^{\circ}$ 
  (from lower to upper) for the Hercules stream stars with the most likely velocity 
  of ($V_R$, $V_{\phi}$)=(32 km s$^{-1}$,185 km s$^{-1}$) at any radii. 
   The broken curves in the middle panel show plots of GSR velocities at $l=20$, $30$ and $40^{\circ}$ 
  (from lower to upper)  in the model with the velocity field ($V_R$, $V_{\phi}$)=(75 km s$^{-1}$, 122 km s$^{-1}$) 
  for short-period mira. The broken curves in the  last panel show the case by the velocity field calculated using a damping bar potential
  (see in the text). 
}\label{fig: fig8}
\end{figure}

Figure 8 shows a comparison of observed radial velocities with the model velocities. 
It shows a clear trend that $V_{\rm GSR}$ for the observed stars decreases with $K_c$.
The model velocity curve should have this tendency. 
Broken curves in each panel indicate the velocity variation with distance
at $l=20$, 30 and $40^{\circ}$ for three different velocity fields: 
the Hercules-stream (top panel),
a short-period blue miras (middle), and a weak-bar model (last).

Here, the Hercules stream has a spatial velocity vector 
($U$, $V$, $W$)$_{\rm LSR}$=($-31.8$, $-35.3$, $-0.8$) in unit of km s$^{-1}$
in the Local Standard of Rest frame \citep{fam05},
where the U, V, and W axes are taken toward the Galactic center, 
toward the direction of Galactic rotation, and toward the Galactic north pole, respectively.
Note that the spatial motion of the Hercules stream is known only in the solar neighborhood, 
and the extension of this stream is not known.
Therefore we assumed for simplicity that the spatial velocities,
$V_R$ and $V_{\phi}$ (rotational and outward motions), are kept at any radii in the Galactic disk. 
The brightest 3 stars ($K_c<2$) in the top panel \ fall in the area
 between two broken curves of $l=30^{\circ}$ and 40$^{\circ}$. Therefore, these stars can be 
associated with Hercules moving group. However, the other fainter stars with $K_c>2$
cannot belong to the same stream, if the given velocity field of the Hercules 
stream is extended to large distances. 
The Hipparcos catalog gives parallax and proper motion data 
for 3 stars in table 4; two of them are in a brightest star group. 
Calculated ($U$, $V$, $W$) velocity components for these two stars are compatible
with the Hercules stream velocity components,   
though one star has negative parallax so that we assumed the distance of 300pc
from the corrected K magnitude.

Middle panel of figure 8 shows a model fitting with a velocity field with $V_R=75$ and $V_{\phi}=122$ km s$^{-1}$,
which have been proposed by \cite{fea00} for short-period blue miras in the solar neighborhood.
This velocity field seems to fit the stars with $K_c>2$ and $l<30^{\circ}$,
though it is not enough for the brightest stars 
with $K_c<3$. \citet{fea00} explained the large deviation of motion of short-period miras 
from the Galactic rotation by oval orbits with large eccentricity produced near the outer Lindblad resonance (OLR),
and suggested that their sample is a mixture of the stars with variety of orbital parameters.

Last panel of figure 8 shows a fit by the velocity field influenced by a Bulge bar. 
The velocity fields are calculated on the basis of a weak-bar linear theory. 
 A simple logarithmic gravitational potential with a few percent deformation due to bulge bar is utilized
in the calculation; see equation (3-77) in \citet{bin87}.
The theory gives a stellar orbit as a sum of two motions in a rotating frame: 
an epicyclic motion, and an oscillating motion produced by a periodic force due to a Bulge bar. The latter is regarded as
 a velocity field in the Galaxy uniquely determined only by the bar gravitational potential, but the former
is determined by the initial conditions of stars and therefore includes arbitrarily randomness.
We calculated the velocity field produced by the bar (the latter), 
and plotted this in the last panel of figure 8.
In order to avoid too large deviations from the equilibrium position due to the Lindblad and corotation resonances,
we introduced a damping constant of the bar potential for convenience, 
and make the velocity field calculable at any radii in the Galaxy  (see Appendix 3 in detail). 
The parameters used in our model is summarized in Table 5 [see details of the parameters 
in the more elaborate calculations of \citet{hab06}].

The last panel of Figure 8 indicates that  the most distant, most deviant stars 
(objects in the lower-right part of the panel), 
can be fit by the parameter ranges within a standard bar model. 
A schematic diagram (shown in figure 9) well explains why such a large deviation from the Galactic rotation  occurs. 
The periodic orbit produced by the Bulge bar (in a rotational coordinate) 
is indicated by the thick ellipse between outer Lindblad and corotation radii. The star is located
at the perigalacticon (small filled circle on the ellipse) when the major axis of the bar passes the guiding center 
of the rotating frame. Beyond the corotation radius, the bar pattern speed ($\sim 60$ km s$^{-1}$ kpc$^{-1}$) 
is faster than the circular rotational speeds of the stars in a standard model of the Galaxy bar;
the periodic orbit is always retrograde. The star moves to the fourth quadrant of the ellipse
(indicated by unfilled circle of figure 9) if the star concerned is located
within 45 degree after the bar passage. Taking the effect of prograde rotation of the
frame into account,  the motion on the rest frame (in GSR) results the star motion toward the Sun. The magnitude of
the velocity toward the Sun depends on the location of the stars in the Galaxy; 
the separation of the star from the bar major axis and 
the separation from the corotation and OLR. 
		
The weak bar theory provides a reasonable explanation for the large negative velocities 
of the distant stars with $K_c=3$--5. However, we found that the parameter values of the standard model 
does not give any good fits for brighter stars, which are located outside of the
OLR. On the periodic orbit (thin ellipse in figure 11), the star comes at apogalacticon
at the bar major-axis passage, and moves to the second quadrant in the ellipse.
The rotational correction to the rest frame gives the star motion receding from the Sun.
Because this is an opposite sense to the observation,  we cannot get any good fit to the velocities of 
bright stars and Hercules group of stars by this model. \citet{deh00} successfully explained 
the motion of Hercules group of stars with the bar model, though he assumed a slightly larger radius of the OLR (7.2 kpc).
The OLR radius, which is close to the solar circular radius,  makes line-of-sight changes of objects
at the solar neighborhood dramatically, and the shear can produce a bimodal velocity distribution function
as made in his model. In our calculation, we have neglected the effect of epicyclic motion 
which acts as a random motion, and introduced a suppression of resonances by decaying bar potential.
The smaller radius of OLR and the damping assumption makes the computation simpler, 
but may lose a strictness of calculations near the OLR.  
In addition, the stellar orbits with large eccentricities are not involved in the linearized theory.
In summary, the velocity field calculated on a basis of the weak-bar theory can give
a good physical insight, and explain the observed stellar velocities
for the distant stars reasonably well, though it is not for the nearby stars. Furthermore,
it is possible that the sampling of nearest stars in the present sample is considerably biased
due to small number.

\begin{figure*}
  \begin{center}
    \FigureFile(90mm,80mm){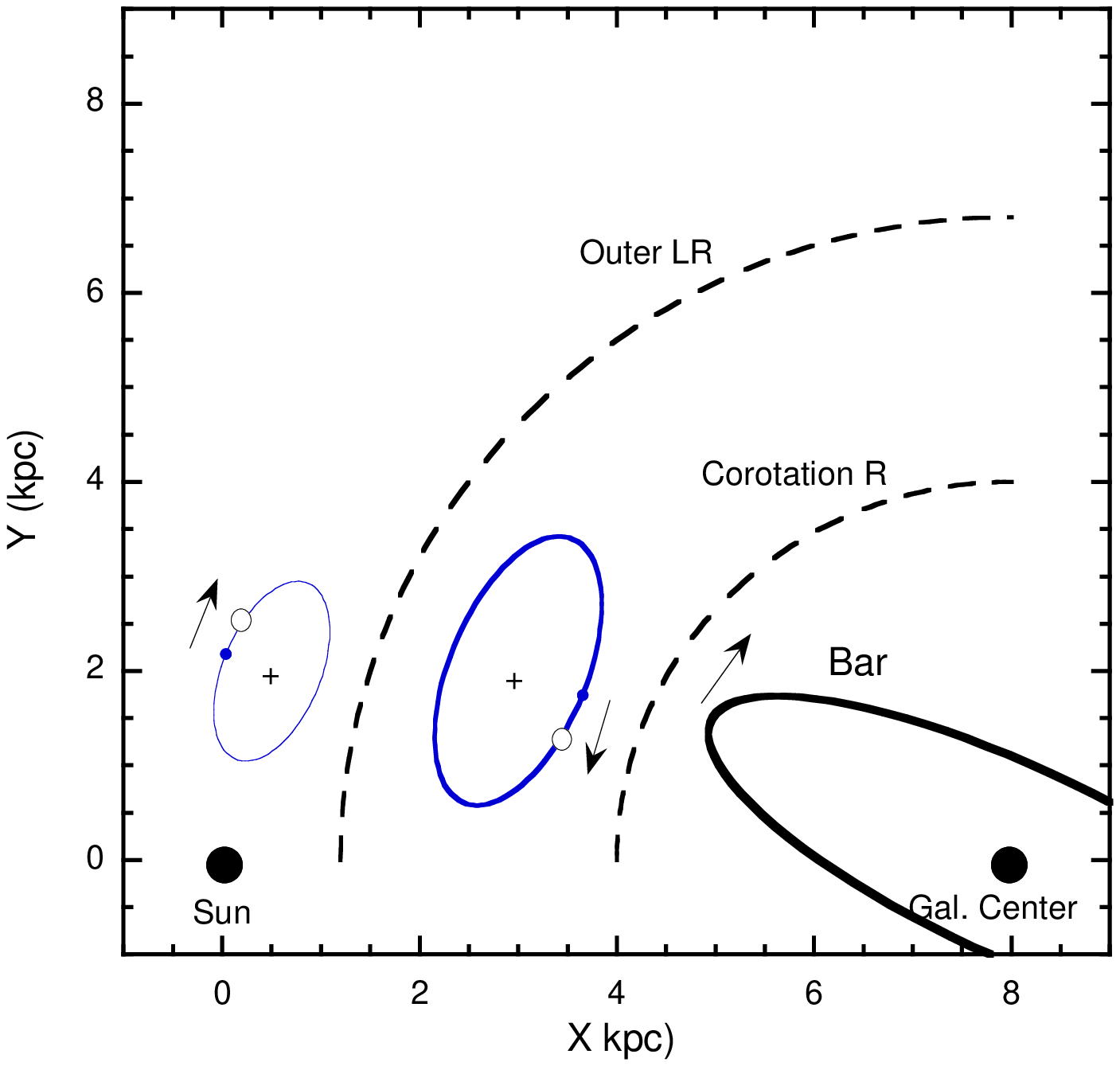}
  \end{center}
  \caption{Schematic diagram of the first quadrant of the Galaxy.
  The Sun is taken as the origin of the X and Y coordinates (indicated by large filled circle), 
  and the X and Y axes toward the Galactic center and the direction of $l=90^{\circ}$.
  The thick line shows the Galactic Bar.
  Locations of the outer Lindblad and corotation resonances are indicated by broken lines.
  The ellipses shown in thin line indicate the stellar orbit seen in the rotating frame.
  The star comes on the minor axis (shown by small filled circle) when the bar major axis
  points toward the center of ellipse (indicated by cross), and it moves at the position shown by
  unfilled circle. Note that, when the guiding center crosses the resonance, 
  the starting position changes their phase by 180$^{\circ}$ on the ellipse, 
  and the direction of rotation changes.  
}\label{fig: fig9}
\end{figure*}

\subsection{Motion perpendicular to the Galactic plane} 
Our data shows that the deviant group of 20 selected stars spreads 
in latitude up to $|b|\sim 30^{\circ}$,
and the average height ($|z|$) is 0.5 kpc. 
These stars may be considered as members of the thick disk, 
which belong to older generations than members of thin disk.
\citet{bin81} considered the resonant excitation of star motion perpendicular
to the Galactic plane due to periodic perturbation by the bar gravitational potential. 
This theory seems to explain well the observed spread
of deviant candidate stars in latitude, at least qualitatively.

A basic equation describing the motion perpendicular to the Galactic plane is written as
[equation (9) of \citet{bin81}]
\begin{equation}    
\ddot{z}+z \{ \nu _0^2 +2 q'_A cos(\kappa t + \phi _0) + 2 q'_B cos[2t(\Omega-\Omega _0)] \}=0 ,
\end{equation}	
where $q'_A$ and $q'_B$ are constants determined by the orbital parameters of the epicyclic 
and forced motion on the Galactic plane, respectively. Here, $\nu _0$ is a basic frequency of oscillation
in z-direction, which is determined by the gravitational potential, and $\Omega$ and $\Omega _0$ are
 the angular speed (a circular rotation) at the guiding center, and the bar pattern speed.
Let us parametrize the constant $\nu _0$ as 
\begin{equation}
\nu _0 =u \Omega .
\label{uomega}
\end{equation}
Here the parameter $u$ describe the flatness of gravitational potential
($0<u<1$).  Because the value of $u$ is not well known for the case of our Galaxy, 
we assume $u=0.5$. The value does not influence strongly on the later discussion. 

The resonance excitation occurs when 
\begin{eqnarray}
 \nu _0 = n \kappa /2,  \hspace{0.5cm} {\rm or} \\
 \nu _0 = n (\Omega  -\Omega _0) ,
 \end{eqnarray}
where n is an integer. Using equation (4), above conditions can be rewritten 
for the Galaxy model with a flat-rotation-curve as
\begin{eqnarray}
   u = n/ \sqrt{2}, \hspace{1.3cm} or \\
  \Omega =  \Omega _0 /[1+ u /(2 n)] .
\end{eqnarray}
Because the first condition is satisfied only for a special potential parameter,
we can neglect the z resonance on the epicyclic motion in our Galaxy.
Because $\Omega  < \Omega _0 $ beyond the corotation radius,
the second condition can be satisfied at various $n$ near the corotation radius,
for example, at radii 1.25 , 1.13, etc. times corotation radius.
These resonant radii are approximately 4--5 kpc in our Galaxy.

The star develops large oscillations perpendicular to the equatorial plane
in a time scale of 10 rotational periods due to resonant coupling \citep{bin81}.
Therefore, the observed displacement of maser sources from the Galactic plane 
seem to be consistent with the resonant coupling theory. 

\begin{figure}
  \begin{center}
    \FigureFile(50mm,50mm){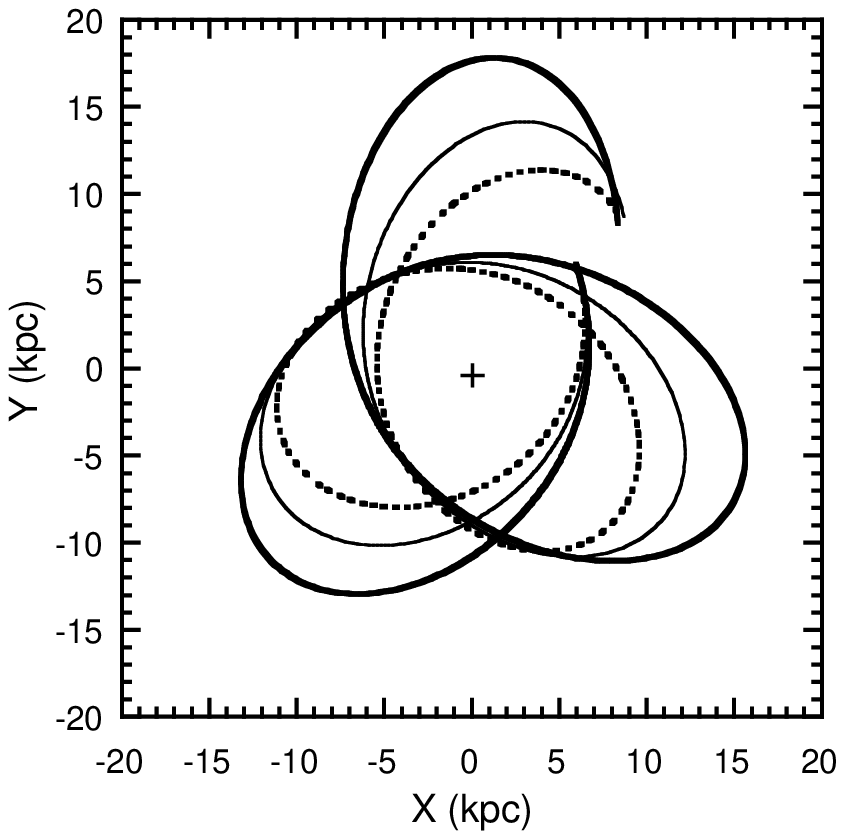}
  \end{center}
  \caption{Stellar orbits in a logarithmic potential (a face-on view to the orbital plane) in a rest frame. 
  Thick, thin, dotted lines indicate stellar orbits with different angular momenta.
  The cross mark shows the center of the model galaxy. 
Unlike $1/r$ potential, the orbit has a smaller curvature inside, 
  but a strong deflection at outer part. 
}\label{fig: fig10}
\end{figure}

\subsection{Further considerations}
Though it is less likely, we investigated a possibility that these deviant stars are a part of a tidal stream 
of a disrupting dwarf galaxy. If it is a relic stream, these stars must be aligned
on a single locus orbiting our Galaxy. We have tried to fit the star positions and 
their radial velocities with a single orbit by a model gravitational potential with a flat rotation curve 
(without influence of a Bulge bar).
However, we could not obtain any reasonable fits which satisfy both positions and radial velocities
within allowable uncertainties in distance. The main reason is that the observational tendency of decreasing radial velocity
with distance between $\sim 0.3$--6 kpc (as shown in figure 8) is hard to realize with any orbits 
in the assumed gravitational potential [$\sim log(r)$]. Unlike the $1/r$ potential,
the orbital locus becomes more or less straight for stars near the corotation radius (see Figure 10)
when the orbit is eccentric in the potential with a flat rotation curve.
To have very small $v_{\rm GSR}$ near corotation radius ($\sim$ 4 kpc) as observed,
the star must move in the direction perpendicular to the line of sight. 
To make a continuos locus of orbit from outer side to inside, the orbit must have a large curvature
at the inside, which cannot make in the assumed potential.

Moreover, the deviant stars have infrared properties similar to those of usual disk SiO maser stars.
Therefore, we deduce that the age and mass distributions of the deviant group 
are not very different from those of usual disk maser stars. 	  
The ages of these maser stars, which are at the AGB phase of stellar evolution, are deduced to be roughly a few Giga yr
(for an initial mass of 1.5--2 $M_{\odot}$; \cite{fea09}). The stars experience the periodic variation
 of gravitational potential by the Bulge bar typically 10 times after their birth ($\sim 2$ Gyr). 
 This time scale is enough for these stars to be deviant from the Galactic rotation.
Possibly, a short-period blue miras are more aged than average maser/infrared stars, so that
the periodic oscillation of the Bulge bar potential appears to affect their motions more severely.

Though we have not completely consume all the possible orbits of tidal streams, 
we think that the deviant group of stars at various distances must comprise of different orbits,
which may be a result of periodic perturbation by the Bulge bar.  
Therefore, all of these considerations support that the origin of the deviant motions
of maser sources is a periodic perturbation of gravitational potential due to the Bulge bar. 

\section{Conclusion}
We detected 84 out of 134 infrared objects off the Galactic plane 
by the SiO $J=1$--0 $v=1$ or 2 lines. Some of these objects 
exhibit large negative radial velocities particularly at $l=20$ -- $40^{\circ}$, 
where the Galactic rotation should give positive ones. 
Their distribution is scattered in the latitude range $\Delta b \sim 60^{\circ}$.
This negative velocity group of stars spreads between 0.3 kpc to 6 kpc in distance. 
It is possible to interpret that the brightest part of this deviant group is the
Hercules stream of stars found in the solar neighborhood, and slightly distant part of this group
as a part of outward flow found in short-period mira, both of which have been explained by 
the resonance effect of the Bulge bar. Though our simple calculation of the velocity field 
based on weak bar theory cannot fit the velocities of the nearest group of selected stars,
it successfully explains the large negative-velocity stars located between the outer Lindblad and corotation resonances.
We have also shown that the resonant coupling due to the periodic perturbation of the Bulge bar
can create the star motion perpendicular to the Galactic plane near the corotation resonance.  
These facts strongly suggest that the deviant group of stars is
produced by the gravitational perturbation of the Bulge bar.  

\

We thank Dr. Tsuyoshi Sakamoto for reading the manuscripts and useful comments.
This research made use of the SIMBAD and VizieR databases operated at CDS, 
Strasbourg, France, and as well as use of data products from 
Two Micron All Sky Survey, which is a joint
project of the University of Massachusetts and Infrared Processing 
and Analysis Center/California Institute of Technology, 
funded by the National Aeronautics and Space Administration and
National Science foundation.

\begin{figure*}
  \begin{center}
    \FigureFile(170mm,70mm){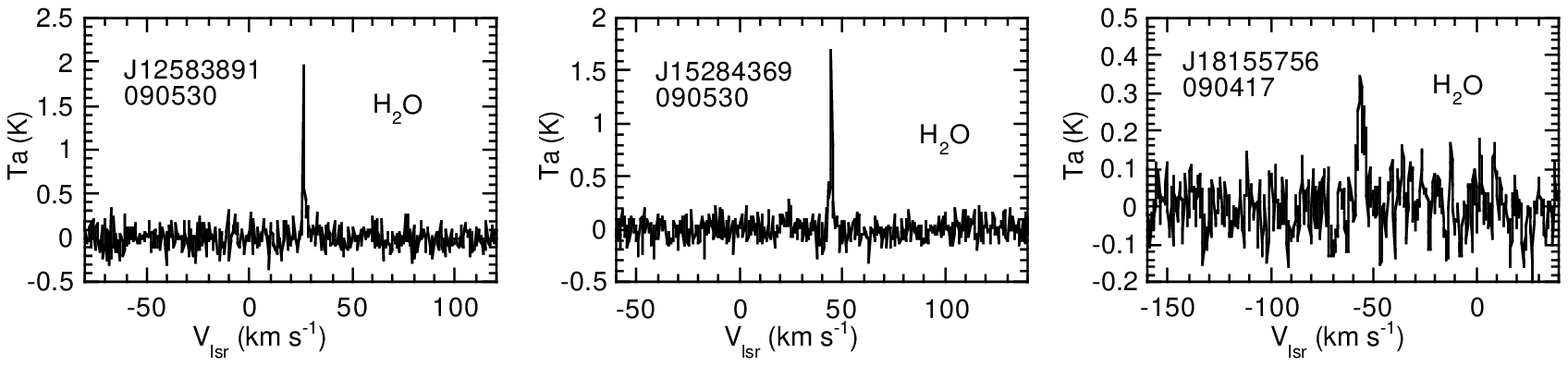}
  \end{center}
  \caption{H$_2$O maser spectra for the detected sources.
}\label{fig: fig11}
\end{figure*}

\section*{Appendix. 1. Individually interesting objects}

\begin{itemize}
\item $J$12370691-1731319  (=T Crv): 
This is an M6  mira with a period 389d \citep{wil04} at high Galactic latitude ($b=45^{\circ}$).  
A previous water maser search gives a negative result \citep{lew97}. We detected a strong
SiO $J=1$-0 $v=1$ line but with no noticeable $v=2$ emission. The very low $v=2/v=1$ intensity ratio
and blue IRAS color ($C_{12}=-0.37$) of this star
fit well with the correlation of this ratio with the IRAS color which was found by \citet{nak07}. 
\citet{vol91} classified the IRAS LRS spectrum of this object as featureless ("F").

\item $J$12583891+2308215 (=T Com):
A previous SiO maser search for this object was negative \citep{nym86}.
OH masers have been detected by \citet{ngu79} at $v_{\rm LSR}=9$ and 23 km s$^{-1}$.
A later observation \citep{che93} found the 1612 MHz emission at $v_{\rm LSR}=23.2$ and 34.1 km s$^{-1}$,
giving the average stellar velocity 28.7 km s$^{-1}$,
which is consistent with the SiO radial velocity at 27.8 km s$^{-1}$ in this paper.
H$_2$O masers have been detected by \citet{kle78} at 25.3 and 27.6 km s$^{-1}$,

\item $J$15591138+1939570 (=V336 Ser): 
This star exhibits featureless IRAS LRS spectrum (LRS class 13), and
\citet{gul97} classified this object as an M-type star. The detection of SiO masers of this star
secures the oxygen richness of this star.

\item $J$18205487+5031432 (=EO Dra = IRAS 18196+5030): 
This is an M7 star at high declination. The IRAS LSR spectrum of this star shows a strong 10 $\mu$m silicate emission
giving an LRS class of 26. \citet{sha95} made spectroscopic observation of this star
deriving the radial velocity of $v_{\rm Helio}\sim -17$ km s$^{-1}$ from the TiO band profile,
which agrees well with the SiO maser velocity $v_{\rm LSR}=-1$ km s$^{-1}$ in this paper.  
No reference of the previous radio observation was found for this star.

\item $J$18213513+8238388 (IRAS 18276+8236): 
This is a relatively bright infrared source located closely to the celestial north pole. 
This objects exhibits strong 10 $\mu$m silicate emission (LSR class 29; \cite{oln86}).  
\citet{coh77} identified this IR object (AFGL 2171) to an M7III star. 
Radio searches for molecular lines had been negative \citep{zuc78,din79,nym92}.
We detected SiO masers at $v_{\rm LSR}=-27$ km s$^{-1}$ for the first time.

\item  $J$19233466+0037583 (V850 Aql =IRAS 19210+0032):   
This s a D-type symbiotic star \citep{phi07} with H$\alpha$ emission \citep{all74}. 
This star was originally misclassified as a planetary nebula, but was corrected later \citep{sab86,ack87}.
Searches for radio continuum emission at 5 and 14 GHz were made with negative results \citep{aaq90}.
No OH or H$_2$O maser search was made. We detected SiO masers in this star for the first time.

\item  $J$19340281+0926061 (IRAS 19316+0919):  
\citet{eng96} detected H$_2$O masers at $V_{\rm LSR}=78.5$ km s$^{-1}$.
OH 1612 MHz maser was a single peak detection 
at $v_{\rm LSR}=76.4$ km s$^{-1}$,
and OH main lines were not detected  \citep{lew97}.
We detected SiO masers at $V_{\rm LSR}=92$ km s$^{-1}$, establishing an accurate
stellar velocity for this object.

\end{itemize}

\section*{Appendix. 2. Water maser observations}
We also observed a few objects by the 22.235 GHz H$_2$O maser lines with the Nobeyama 45m telescope 
during the same period of SiO observations as a backup for bad weather condition. 
Though the H$_2$O maser observations are limited,  interesting objects are involved.
The HEMT 22GHz receivers was used for observations and 
the conversion factor of antenna temperature to flux density is about 3.0 Jy/K.
We have detected 3 objects. The line parameters of the H$_2$O masers are given in table 6,
and the line profiles are given in figure 11. 




\onecolumn
\section*{Appendix. 3. Theory of star motion in a weak bar potential}
In a frame rotating with angular speed $\Omega$,
two-dimensional equations of motion for a test particle moving in the Galactic plane 
 are written  as
\begin{eqnarray}
 \ddot x\ = 2 \Omega  \dot y\ -\ \partial \Phi /\partial x \ +\  \Omega^2 x   \\
 \ddot y\ =-2 \Omega   \dot x\ -\ \partial \Phi /\partial y \ +\  \Omega^2 y
\end{eqnarray}
where $\Phi $ is a gravitational potential (see \cite{bin87}) and the origin of the coordinates is taken at the center of the Galaxy. 
For a weak bar-like potential with $m$-fold symmetry, 
the gravitational potential can be written as
\begin{equation}
\Phi = \Phi_0(r) \ +\  \Phi_1(r) {\rm cos}\{ m [\theta - (\Omega_0-\Omega) t] \}
\end{equation}
where $\Omega_0$ is the pattern speed of a bar, and $\theta$ is the angle of the particle position vector from the x-axis in the rotating frame.
In this paper, we only consider a case of a weak bar potential, where the effect due the second term is small compared with the first term, i.e.,  $|\Phi_1(r)|   \ll  |\Phi_0(r)|$, and $m=2$.
The angular rotation speed of the frame is determined by the balance of centrifugal force with the 0th-order gravitational attraction,
\begin{equation}
  \Omega^2 r_0 \ =\  (\partial \Phi_0 /\partial r)|_{x=r_0, y=0}
\end{equation}
Putting small deviations of the particle position from the equilibrium position ($x_0=r_0$, $y_0=0$)
\begin{eqnarray}
x_1=x-x_0 \\
y_1=y-y_0,
\end{eqnarray}
we obtain
\begin{eqnarray}   
 \ddot x_1\ = 2 \Omega  \dot y_1\ +\ x_1 [\Omega^2- \partial^2 \Phi_0 /\partial r^2 |_{x=r_0}] \ -\   \partial \Phi_1 /\partial r |_{x=r_0} {\rm cos}(m\Omega_1 t) \\
 \ddot y_1\ = - 2 \Omega   \dot x_1\  \    -\ \Phi_1|_{x=r_0} m {\rm sin}(m \Omega_1 t)/r_0\   \hspace{4cm}
\end{eqnarray}
in the first order approximation, 
where $ \Omega_1  \equiv \Omega_0 -\Omega$.
For the case of a galaxy with a flat rotation curve, the gravitational potential can be written as
\begin{equation}   
 \Phi_0(r)\ = (v_0^2 /2) {\rm ln} (a^2+x^2+y^2) 
\end{equation}
where $v_0$ is a circular rotation velocity of a particle (constant) and $a$ is a core radius.
The bar potential is often approximated (e.g., \cite{hab06}) as 
\begin{equation}
 \Phi = (v_0^2 /2) {\rm ln} [a^2+x^2 (1-\epsilon)+y^2  (1+\epsilon)].    \\
 \end{equation}
 We set the bar potential $\Phi_1 (r)$ as close as to the first order term in  $\epsilon$ of the above equation,
\begin{equation}
  \Phi_1(r) = - \epsilon v_0^2 r^2/(a^2+r^2 ) .
  \end{equation}
In such an approximation, we finally have equations of motion as
\begin{eqnarray}    
 \ddot x_1\ = 2 \Omega  \dot y_1\ +\  (4\Omega^2 -\kappa^2)  x_1 \ +\  B_x {\rm cos}(m \Omega_1 t) \\
 \ddot y_1\ = - 2 \Omega   \dot x_1\  + \   B_y {\rm sin }(m \Omega_1 t) ,
\end{eqnarray}
where
\begin{eqnarray}    
 B_x= -\partial \Phi_1 /\partial r |_{x=r_0} \ = \  2 \epsilon v_0^2  r_0\ a^2/(a^2+r_0^2)^2 \\
 B_y= - m \Phi_1 |_{x=r_0} /r_0 \ = \ m \epsilon v_0^2 r_0  / (a^2+r_0^2) ,
\end{eqnarray}
and
\begin{equation}  
 \kappa^2\ \equiv\ 3\Omega^2 + \partial^2 \Phi_0 /\partial r^2 |_{x=r_0}   = \ 2 \Omega^2  \\
\end{equation}
In the case of decaying bar potential, we use
\begin{equation}     
 \epsilon =  \epsilon_0 \ {\rm exp}(- \gamma t) ,
\end{equation}
where  $\epsilon_0$ is a constant.
Note that the equations of motion (20) and (21)  are linear with respect to $x_1$ and $y_1$ with additional forced oscillating terms. 
These equations  have a general solution of linear combination of
epicyclic and forced-oscillation terms, 
\begin{eqnarray}   
 x_1= A_x {\rm cos}( \kappa t +\phi)  + E_x (t) {\rm exp}(-\gamma t)\\
 y_1= A_y {\rm sin} (\kappa t +\phi)  + E_y(t) {\rm exp}(-\gamma t),
\end{eqnarray}
where the epicyclic terms must satisfy the following condition
\begin{equation}    
 A_y =2 (\Omega/\kappa) A_x ,
 \end{equation}
 and $A_x$  (or  $A_y$) and $\phi$ are arbitrary constants determined by the initial condition, and
 the forced oscillation terms are expressed as
 \begin{eqnarray}  
E_x (t)= [(N_{xcx}  B_x + N_{xcy} B_y) {\rm cos}(m \Omega_1 t) + [(N_{xsx} B_x + N_{xsy} B_y) {\rm sin}(m \Omega_1 t)]/d_{xy}  \\
E_y (t)= [(N_{ycx}  B_x + N_{ycy} B_y) {\rm cos}(m \Omega_1 t) + [(N_{ysx} B_x + N_{ysy} B_y) {\rm sin}(m \Omega_1 t)]/d_{xy}.
 \end{eqnarray}
 Here,  the denominator is calculated as
  \begin{equation}
d_{xy}  = [ (m \Omega_1- \kappa)^2 + \gamma^2] [ (m \Omega_1+\kappa)^2+\gamma^2]  (m^2 \Omega_1^2+\gamma^2)^2  ,
 \end{equation}
 and the terms in numerators are written as
 \begin{eqnarray}      
N_{xcx} =   -(m^2 \Omega_1^2+ \gamma^2)^2 (m^2 \Omega_1^2-\kappa^2- \gamma^2)  ,   \\
N_{xcy}=     2 m \Omega \Omega_1 (m^2 \Omega_1^2+ \gamma^2) (m^2 \Omega_1^2-\kappa^2-3  \gamma^2)  , \\
N_{xsx} =    -2  \gamma m \Omega_1 (m^2 \Omega_1^2+ \gamma^2)^2  ,  \\
N_{xsy}=     2  \gamma \Omega (m^2 \Omega_1^2+ \gamma^2) (3 m^2 \Omega_1^2-\kappa^2- \gamma^2) , \\
N_{ycx} =    -2  \gamma \Omega (m^2 \Omega_1^2+ \gamma^2) (3 m^2 \Omega_1^2-\kappa^2- \gamma^2)    \\
N_{ycy}=     2  \gamma m \Omega_1 (m^4 \Omega_1^4+8 m^2 \Omega^2 \Omega_1^2-2\kappa^2 m^2 \Omega_1^2   \nonumber  \\
             +2  \gamma^2 m^2 \Omega_1^2-4\kappa^2 \Omega^2-8  \gamma^2 \Omega^2+\kappa^4+2  \gamma^2\kappa^2+ \gamma^4) , \\
N_{ysx} =    2 m \Omega \Omega_1 (m^2 \Omega_1^2+ \gamma^2) (m^2 \Omega_1^2-\kappa^2-3  \gamma^2)  ,  \\
N_{ysy}=     -m^6 \Omega_1^6+(2\kappa^2- \gamma^2-4 \Omega^2) \ m^4 \Omega_1^4 \nonumber \\
              +[(4\kappa^2+24  \gamma^2) \Omega^2 -\kappa^4-4  \gamma^2\kappa^2+ \gamma^4] \ m^2 \Omega_1^2  \nonumber\\
			+(-4  \gamma^2\kappa^2-4  \gamma^4) \Omega^2+ \gamma^2\kappa^4+2  \gamma^4\kappa^2+ \gamma^6    .
 \end{eqnarray} 
The denominator, $d_{xy}$, is always positive at the Lindblad-  and corotation-resonance radii 
(i.e.,  $m^2\Omega_1^2 -\kappa^2=0$ and $\Omega_1=0$),
when the damping term ($|\gamma| >0 $) is introduced. Therefore stellar orbits are calculable at any radii
as far as the deviation from the equilibrium position is small. Note that the cross terms, 
$N_{xsx}$,  $N_{xsy}$, $N_{ycx}$, and $N_{ycy}$ are proportional to $\gamma$. When $|\gamma|$ is much smaller than $|\Omega_1|$,
these terms are negligible and the major axis of the elliptic orbit is oriented perpendicularly to the radial direction (as shown in Figure 9).
However, near the resonance, these terms influence to the orientation of the elliptic orbit 
such as the major axis of the orbit is slightly inclined toward
the radial direction of the Galaxy. This effect  produces larger observed radial velocities of stars near the resonant position.
It is well known in the  limit of $\gamma=0$ that the amplitude terms  in equations (29) and (30) change sign when the equilibrium position
crosses the Lindblad-resonances.  Therefore, the phase of the particle 
 in the elliptic orbit due to forced oscillation varies by 180 degree at the resonance (see Figure 9). However,
this sudden change is moderated in the damping model, causing a cycling shape of the curves 
 near the outer Lindblad resonance as shown in the last panel of Figure 8.
	 
It is believed that  bars in the gas-rich spirals are short lived (e.g., \cite{bou05}). 
Furthermore, the effect of a triaxial halo, or a central massive blackhole
may also destroy the bar within a Hubble time scale \citep{ide00,hoz05}.
Therefore, the decaying (or growing) bar model presented in this paper well facilitates to investigate stellar orbits 
near resonances in the Galaxy. 

\clearpage
\tabcolsep 1pt
\begin{longtable}{crrrrrrrrr}
  \caption{Observational results of SiO Masers.} 
\hline  \hline
  & \multicolumn{4}{c}{\underline{  SiO $J=1$--0 $v=1$ line  }} & \multicolumn{4}{c}{\underline{  SiO $J=1$--0 $v=2$ line  }} & \\
 2MASS name &  $Ta$  &   $V_{\rm lsr}$  & L.F.  &  $rms$  &      $Ta$  &    $V_{\rm lsr}$ & L.F.  &  $rms$ & obs. date \\  
   & {\tiny (K)} & {\tiny (km s$^{-1}$)} & {\tiny (K km s$^{-1}$)} &  {\tiny (K)} & {\tiny (K)} & {\tiny (km s$^{-1}$)} & {\tiny (K km s$^{-1}$)} &  {\tiny (K)} & {\tiny (yymmdd.d)}\\
\hline
\endfirsthead
\hline \hline
   & \multicolumn{4}{c}{\underline{  SiO $J=1$--0 $v=1$ line  }}  & \multicolumn{4}{c}{\underline{  SiO $J=1$--0 $v=2$ line  }} &  \\
2MASS name &  $Ta$  &  $V_{\rm lsr}$  & L.F.  &  $rms$  &      $Ta$  &    $V_{\rm lsr}$ & L.F.  &  $rms$ & obs. date \\  
  & {\tiny (K)} & {\tiny (km s$^{-1}$)} & {\tiny (K km s$^{-1}$)} &  {\tiny (K)} & {\tiny (K)} & {\tiny (km s$^{-1}$)} & {\tiny (K km s$^{-1}$)} &  {\tiny (K)} & {\tiny (yymmdd.d)}\\
\hline
\endhead
\hline
\endfoot
\hline
\multicolumn{10}{l}{Table note: "$Ta$" is the antenna temperature at the intensity peak.}\\ 
\multicolumn{10}{l}{  "$V_{\rm lsr}$" is the radial velocity in the Local Standard of Rest frame.} \\
\multicolumn{10}{l}{ "L.F." is the integrated line intensity.}  \\
\multicolumn{10}{l}{ "$rms$" is the root mean square of the noise level.}\\
\endlastfoot
$J12370691-1731319$ &   1.473  &$-$33.5 &   3.113  &  0.051  &   ...   &   ...  &    ...   &  0.041  &  090531\\
$J12583891+2308215$ &   2.078  &   27.8 &   5.549  &  0.055  &  1.264  &   27.8 &   4.323  &  0.046  &  090531\\
$J15284369+0349430$ &   6.648  &   44.2 &  11.289  &  0.082  &  4.822  &   44.2 &  12.526  &  0.071  &  090529\\
$J15591138+1939570$ &   1.017  &   14.6 &   2.618  &  0.078  &  0.707  &   14.7 &   2.191  &  0.067  &  090531\\
$J16122976+2453570$ &   1.651  &$-$16.8 &   5.091  &  0.083  &  1.401  &$-$15.6 &   3.902  &  0.066  &  090531\\
$J16292643-1920509$ &   1.259  & $-$9.2 &   4.769  &  0.084  &  0.482  & $-$7.0 &   2.384  &  0.073  &  090529\\
$J16510590+1020515$ &   4.213  &   19.4 &  17.956  &  0.103  &  2.992  &   19.4 &  10.998  &  0.094  &  090531\\
$J16524821+0524269$ &   2.096  &$-$46.3 &   9.523  &  0.058  &  0.668  &$-$52.4 &   3.231  &  0.050  &  090529\\
$J16534478+4857022$ &   3.951  &   16.8 &  11.554  &  0.129  &  2.703  &   16.6 &   8.947  &  0.104  &  090531\\
$J17210403+2655505$ &   3.114  &   25.8 &  14.033  &  0.178  &  0.892  &   25.8 &   0.814  &  0.161  &  090530\\
$J17312879+3229525$ &   1.278  &$-$24.8 &   7.159  &  0.112  &  1.160  &$-$24.8 &   3.926  &  0.998  &  090530\\
$J17331391+0820390$ &   0.467  &    8.1 &   1.243  &  0.132  &  0.655  &    8.0 &   1.167  &  0.109  &  090417\\
$J17364445+1051070$ &   1.781  &$-$55.3 &   5.434  &  0.208  &  2.074  &$-$55.4 &   5.241  &  0.175  &  090530\\
$J18000391+2335371$ &   1.310  &    3.3 &   3.068  &  0.123  &  0.727  &    3.8 &   0.955  &  0.105  &  090530\\
$J18024911-0632355$ &   0.217  &$-$39.1 &   0.290  &  0.062  &  0.267  &$-$38.6 &   0.506  &  0.049  &  090601\\
$J18080907-0649058$ &   1.097  &$-$50.4 &   2.094  &  0.227  &  1.147  &$-$50.4 &   2.294  &  0.186  &  090417\\
$J18102890-0237427$ &   0.678  &   65.3 &   1.819  &  0.080  &  0.494  &   66.0 &   2.652  &  0.080  &  090419\\
$J18105856+0753085$ &    ...   &   ...  &    ...   &  0.103  &  0.561  &$-$64.1 &   1.217  &  0.082  &  090530\\
$J18110144-0142340$ &   0.440  &   63.1 &   2.012  &  0.080  &  0.376  &   61.4 &   1.265  &  0.070  &  090531\\
$J18120477-0607247$ &   0.265  &  120.7 &   0.595  &  0.043  &  0.247  &  120.5 &   0.767  &  0.038  &  040529\\ 
$J18155756+0120106$ &   0.543  &   21.3 &   3.232  &  0.089  &  0.358  &   21.6 &   2.483  &  0.068  &  090417\\
$J18162782-0425247$ &   2.177  &   57.7 &   4.115  &  0.288  &  2.086  &   58.6 &   3.711  &  0.239  &  090417\\
$J18170265-0720564$ &   0.434  &$-$10.8 &   1.432  &  0.073  &  0.338  &$-$10.9 &   1.344  &  0.066  &  090418\\
$J18192465-0439593$ &   0.214  &    1.0 &   1.262  &  0.046  &  0.159  & $-$1.9 &   0.995  &  0.047  &  090418\\
$J18193355+0354498$ &   0.166 &    49.4 &   0.377  &  0.055  &  0.207  &   49.8 &   0.544  &  0.050  &  090418\\ 
$J18203449-0342095$ &   0.805  &   21.4 &   1.192  &  0.089  &  0.674  &   21.2 &   2.025  &  0.085  &  090418\\
$J18205487+5031432$ &   2.184  & $-$0.7 &   5.533  &  0.160  &  1.365  & $-$1.0 &   3.410  &  0.146  &  090531\\
$J18212420+0229022$ &   0.309  &   18.3 &   1.262  &  0.065  &  0.371  &   21.6 &   1.647  &  0.065  &  090418\\
$J18213513+8238388$ &   1.168  &$-$27.0 &   3.210  &  0.123  &  0.583  &$-$26.6 &   1.756  &  0.104  &  090531\\
$J18213854-0355447$ &   0.581  &   71.5 &   3.090  &  0.080  &  0.593  &   72.8 &   2.598  &  0.069  &  090529\\
$J18223249-0305115$ &   0.470  &   55.9 &   1.111  &  0.078  &  0.381  &   55.7 &   0.589  &  0.069  &  090418\\
$J18231986+0929569$ &   0.597  &$-$14.4 &   1.514  &  0.058  &  0.613  &$-$15.3 &   1.174  &  0.056  &  090418\\
$J18250570-0032320$ &   0.517  &   99.1 &   1.762  &  0.069  &  0.381  &   99.1 &   1.896  &  0.064  &  090418\\
$J18253335+0856472$ &   0.326  & $-$7.6 &   0.900  &  0.052  &  0.200  & $-$7.6 &   0.902  &  0.045  &  090529\\
$J18264298-0024486$ &   0.282  &  125.0 &   0.876  &  0.056  &  0.192  &  124.8 &   0.839  &  0.048  &  090531\\
$J18274403+0025128$ &   0.962  &   99.6 &   2.578  &  0.153  &  0.833  &  100.1 &   2.022  &  0.118  &  090417\\
$J18302847+0523383$ &   0.712  &   28.3 &   2.580  &  0.124  &  0.595  &   28.4 &   1.606  &  0.096  &  090417\\
$J18331997+0425410$ &   0.980  &   73.1 &   3.135  &  0.093  &  0.643  &   72.2 &   2.872  &  0.077  &  090531\\
$J18354242+0905384$ &   1.006  &   51.4 &   2.662  &  0.072  &  0.895  &   51.3 &   2.440  &  0.068  &  090418\\
$J18364611+0845469$ &   0.989  & $-$9.7 &   2.518  &  0.105  &  0.630  & $-$9.8 &   2.140  &  0.093  &  090531\\
$J18384242+0541298$ &   0.652  &   40.8 &   2.482  &  0.102  &  0.604  &   41.5 &   2.251  &  0.080  &  090531\\
$J18420916+0801180$ &   0.303  &$-$96.4 &   0.564  &  0.057  &  0.410  &$-$97.1 &   1.882  &  0.064  &  090418\\
$J18454767-1148074$ &   0.983  &  102.6 &   1.449  &  0.067  &  0.958  &  102.4 &   2.168  &  0.090  &  090418\\
$J18501116-1007570$ &   0.329  &   55.2 &   1.952  &  0.056  &  0.218  &   46.6 &   1.187  &  0.049  &  090601\\
$J18512520+1202084$ &   0.386  &   35.2 &   1.250  &  0.067  &  0.367  &   36.1 &   1.531  &  0.060  &  090418\\
$J18523817+1733113$ &   0.264  &   97.7 &   1.179  &  0.056  &  0.125  &   95.9 &   0.317  &  0.048  &  090531\\
$J18524262-0951445$ &   0.339  &  108.0 &   1.047  &  0.081  &  ...    &    ... &   ...    &  0.070  &  090601\\
$J18530987-1329244$ &   0.599  &   85.0 &   1.725  &  0.070  &  0.399  &   87.2 &   1.079  &  0.399  &  090417\\
$J18550366-1328438$ &   0.437  &   29.3 &   1.322  &  0.078  &  0.447  &   26.8 &   0.882  &  0.066  &  090601\\
$J18554765-1415207$ &   0.650  &   83.1 &   1.498  &  0.082  &  0.338  &   83.6 &   0.547  &  0.069  &  090601\\
$J18572211-0831208$ &   ...    &    ... &   ...    &  0.084  &  0.320  &   47.6 &   1.020  &  0.081  &  090418\\
$J18572648+1349096$ &   0.566  &   76.2 &   0.523  &  0.130  &  ...    &    ... &   ...    &  0.105  &  090417\\
$J18575917+1413196$ &   0.500  &   86.2 &   1.303  &  0.103  &  0.551  &   86.6 &   0.985  &  0.093  &  090418\\
$J18594368-0924127$ &   0.576  &   39.7 &   1.010  &  0.131  &  0.450  &   41.5 &   0.519  &  0.129  &  090418\\
$J19004752-0742491$ &   0.273  &   84.7 &   0.681  &  0.083  &  ...    &    ... &   ...    &  0.073  &  090419\\
$J19010471-1050004$ &   0.376  &   28.1 &   1.487  &  0.089  &  0.442  &   28.3 &   1.614  &  0.080  &  090419\\
$J19010944+1538566$ &   0.738  &   24.3 &   1.800  &  0.086  &  0.340  &   24.2 &   1.316  &  0.076  &  090530\\
$J19012574-0529398$ &   1.014  &   87.3 &   3.152  &  0.096  &  0.569  &   87.3 &   1.694  &  0.081  &  090419\\
$J19020569-1236483$ &   0.430  &$-$21.4 &   1.058  &  0.068  &  0.432  &$-$20.7 &   1.280  &  0.061  &  090530\\
$J19024407-0611124$ &   0.538  &   95.6 &   0.688  &  0.083  &  0.337  &   95.3 &   0.422  &  0.075  &  090419\\
$J19050061+2310298$ &   0.486  &$-$23.1 &   2.548  &  0.084  &  0.732  &$-$21.0 &   2.427  &  0.077  &  090419\\
$J19054911-1212243$ &   1.253  &$-$16.8 &   3.917  &  0.123  &  0.979  &$-$16.4 &   3.563  &  0.114  &  090419\\
$J19062439-1509534$ &   0.990  &$-$48.7 &   2.833  &  0.114  &  1.088  & $-$48. &   2.997  &  0.099  &  090601\\
$J19074038-0515161$ &   0.415  &   99.0 &   0.574  &  0.063  &  ...    &    ... &   ...    &  0.056  &  090601\\
$J19085920-1510032$ &   1.315  &   46.7 &   2.900  &  0.107  &  0.761  &   46.7 &   1.289  &  0.761  &  090419\\
$J19090454+2939292$ &   3.692  &$-$13.4 &  10.319  &  0.105  &  2.520  &$-$12.8 &   5.419  &  0.091  &  090530\\
$J19094939-0804034$ &   0.333  &   30.0 &   0.991  &  0.059  &  0.251  &   30.8 &   0.678  &  0.056  &  090601\\
$J19111079-0227493$ &   0.481  &   51.2 &   1.379  &  0.066  &  0.213  &   52.1 &   0.675  &  0.055  &  090601\\
$J19211169+0320578$ &   0.283 &$-$21.1 &   0.683  &  0.064  &  0.333  &$-$23.3 &   0.785  &  0.064  &  090419\\
$J19233466+0037583$ &   1.674  &   26.1 &   4.442  &  0.128  &  1.143  &   26.3 &   1.085  &  0.115  &  090419\\
$J19234517+7141137$ &   3.018  &   13.7 &   8.656  &  0.189  &  5.438  &   14.3 &  15.792  &  0.169  &  090531\\
$J19240522-0722442$ &   0.589  &   31.2 &   1.061  &  0.116  &  0.495  &   31.3 &   1.030  &  0.100  &  090531\\
$J19340281+0926061$ &   0.367  &   92.2 &   0.974  &  0.097  &  0.431  &   91.9 &   2.237  &  0.090  &  090419\\
$J19341153+1958285$ &   1.616  &   45.5 &   5.349  &  0.124  &  1.480  &   46.0 &   4.511  &  0.101  &  090531\\
$J19360670+0945041$ &   0.811  &   17.5 &   1.919  &  0.112  &  0.586  &   17.3 &   1.512  &  0.093  &  090531\\
$J19433804+1403158$ &   1.177  &   28.3 &   2.897  &  0.151  &  0.630  &   28.5 &   1.648  &  0.098  &  090601\\
$J19444164+0513039$ &   0.618  &   34.8 &   1.674  &  0.096  &  0.441  &   36.4 &   1.012  &  0.090  &  090419\\
$J19464724+1549014$ &   0.322  &   56.7 &   0.752  &  0.064  &  0.331  &   56.9 &   0.852  &  0.058  &  090601\\
$J19510953+1354375$ &   0.932  &   40.7 &   1.566  &  0.099  &  0.468  &   39.7 &   1.079  &  0.087  &  090419\\
$J19542885+1943430$ &   3.802  &   66.7 &   7.192  &  0.162  &  3.300  &   66.7 &   7.487  &  0.138  &  090419\\
$J20003856+1331331$ &   0.975  &$-$25.8 &   3.096  &  0.166  &  0.489  &$-$25.0 &   1.638  &  0.122  &  090530\\
$J20042163+1748345$ &   0.734  &   28.1 &   1.746  &  0.091  &  0.445  &   27.3 &   1.734  &  0.080  &  090419\\
$J20120916+1116516$ &   0.867  &$-$13.8 &   2.811  &  0.109  &  0.753  &$-$13.9 &   2.735  &  0.096  &  090419\\
$J20172312+1718459$ &   0.614  &    5.2 &   1.101  &  0.093  &  0.566  &    5.0 &   1.228  &  0.092  &  090419\\
$J20292219+1311116$ &   0.973  &   24.1 &   3.790  &  0.107  &  0.102  &   24.2 &   2.699  &  0.071  &  090419\\
\end{longtable}

\clearpage
\begin{longtable}{lrrr}
  \caption{negative results.} 
\hline  \hline
               & SiO v=1  & SiO v=2 &  \\
2MASS name     &  $rms$  &   $rms$ & obs. date \\  
   & {\tiny (K)} &  {\tiny (K)} & {\tiny (yymmdd.d)}\\
\hline
\endfirsthead
\hline \hline
              & SiO v=1  & SiO v=2 &  \\
2MASS name    &  $rms$  &   $rms$ & obs. date \\  
              & {\tiny (K)} &  {\tiny (K)} & {\tiny (yymmdd.d)}\\
\hline
\endhead
\hline
\endfoot
\hline
\multicolumn{4}{l}{Table note: "$rms$" is the root mean square of the noise level.}\\
\endlastfoot
$J16322460-1312013$ &  0.093 &   0.079 & 090531\\
$J16521876-1830343$ &  0.103 &   0.088 & 090531\\
$J17171484-0230186$ &  0.059 &   0.049 & 090529\\
$J17551370+1143462$ &  0.074 &   0.062 & 090529\\
$J18042450-0802252$ &  0.083 &   0.078 & 090419\\
$J18135847-0815296$ &  0.072 &   0.080 & 090418\\
$J18182494-0331378$ &  0.177 &   0.141 & 090531\\
$J18240372+0636258$ &  0.143 &   0.118 & 090417\\
$J18243205-0132065$ &  0.069 &   0.057 & 090529\\
$J18270226-0106095$ &  0.152 &   0.131 & 090530\\
$J18275079+0752206$ &  0.072 &   0.064 & 090418\\
$J18283699+0936501$ &  0.104 &   0.084 & 090531\\
$J18294430+0104534$ &  0.072 &   0.080 & 090418\\
$J18295773+0114056$ &  0.071 &   0.061 & 090530\\
$J18414435+0802164$ &  0.085 &   0.075 & 090531\\
$J18421582+0731314$ &  0.075 &   0.667 & 090418\\
$J18431907+0733137$ &  0.067 &   0.059 & 090418\\
$J18465164-1157048$ &  0.081 &   0.075 & 090418\\
$J18495300-1104202$ &  0.084 &   0.069 & 090601\\
$J18530734-1155310$ &  0.194 &   0.164 & 090417\\  
$J18562524-0744227$ &  0.109 &   0.119 & 090418\\
$J18574985+2031371$ &  0.361 &   0.296 & 090417\\
$J19030151+1656312$ &  0.093 &   0.080 & 090531\\
$J19034074-1409330$ &  0.097 &   0.091 & 090419\\
$J19060940+1738007$ &  0.084 &   0.072 & 090531\\
$J19062416+1739313$ &  0.053 &   0.047 & 090530\\  
$J19102783-1541355$ &  0.066 &   0.056 & 090601\\
$J19105141-0328410$ &  0.119 &   0.099 & 090419\\
$J19134664-0326081$ &  0.118 &   0.107 & 090419\\
$J19141725-0850549$ &  0.074 &   0.064 & 090419\\
$J19145484-0225287$ &  0.104 &   0.087 & 090530\\
$J19173939-1322488$ &  0.082 &   0.071 & 090531\\
$J19181178-0721437$ &  0.071 &   0.061 & 090601\\
$J19182271-0242108$ &  0.116 &   0.102 & 090419\\
$J19222258-1418050$ &  0.100 &   0.098 & 090419\\
$J19265373-0104251$ &  0.110 &   0.100 & 090419\\
$J19410829+0202312$ &  0.084 &   0.084 & 090531\\
$J19433664+1049180$ &  0.118 &   0.102 & 090531\\
$J19483842+2759358$ &  0.076 &   0.060 & 090530\\
$J19492572+0231306$ &  0.115 &   0.093 & 090530\\
$J19552508+0156036$ &  0.121 &   0.098 & 090601\\
$J19553800-0018428$ &  0.122 &   0.105 & 090530\\
$J19554049+1805361$ &  0.089 &   0.076 & 090419\\
$J19595639+1151450$ &  0.128 &   0.111 & 090531\\
$J20030250+0544166$ &  0.091 &   0.076 & 090419\\
$J20075461+1842544$ &  0.089 &   0.084 & 090419\\
$J20080544+1516428$ &  0.105 &   0.088 & 090531\\
$J20571628+0258445$ &  0.056 &   0.060 & 090601\\
\end{longtable}
%
 \tabcolsep 0.5pt
\begin{longtable}{lrrrrrrrrr}
 \caption{Infrared properties of the observed sources.} 
\hline  \hline
2MASS name       & $l$ & $b$  & $K^{*}$ & $H-K$ & $K_c$ & IRAS & $F_{12}$ &  $C_{12}\ $  & $v$(SiO)$^{\sharp}$\\  
                 & ($\circ$) &  ($\circ$)    &   &     &       &      &  (Jy)    &      & {\tiny (km s$^{-1}$)} \\ 
\hline
\endfirsthead
\hline \hline
2MASS name       & $l$ & $b$  & $K^{*}$ & $H-K$ & $K_c$ & IRAS & $F_{12}$ &  $C_{12}\ $  & $v$(SiO)$^{\sharp}$\\  
                 & ($\circ$)&  ($\circ$)    &   &     &       &      &  (Jy)    &      &  {\tiny (km s$^{-1}$)} \\ 
\hline
\endhead
\hline
\endfoot
\hline
\multicolumn{9}{l}{ $^{*}$ The 2MASS photometric uncertainties are about 0.2 mag for bright stars }\\
\multicolumn{9}{l}{  with $K<4$; see,  \it  http://www.ipac.caltech.edu/2mass/releases/allsky/doc/sec2\_2.html  .}\\
\multicolumn{9}{l}{$^{\sharp}$ "$v$(SiO) is the average radial velocity of SiO maser lines.} \\
\endlastfoot
$J12370691-1731319$ & 298.083 & 45.211 & 2.598 & 0.765 & 2.216 & 12345$-$1715 & 34.50 &$-$0.373 &$-$33.5\\ 
$J12583891+2308215$ & 325.571 & 85.690 & 3.270 & 0.736 & 2.930 & 12562+2324 & 33.70 &$-$0.233 & 27.8\\ 
$J15284369+0349430$ & 8.103 & 45.840 & 3.364 & 0.858 & 2.848 & 15262+0400 & 46.20 &$-$0.175 & 44.2\\ 
$J15591138+1939570$ & 33.459 & 46.526 & 3.449 & 0.664 & 3.213 & 15569+1948 & 10.60 &$-$0.433 & 14.6\\ 
$J16122976+2453570$ & 41.919 & 45.074 & 2.789 & 0.700 & 2.501 & 16103+2501 & 17.40 &$-$0.321 &$-$16.2\\ 
$J16292643-1920509$ & 357.592 & 19.667 & 2.381 & 0.689 & 2.109 & 16265$-$1914 & 28.90 &$-$0.385 &$-$8.1\\ 
$J16322460-1312013$ & 3.227 & 22.945 & 4.132 & 0.604 & 3.982 & 16296$-$1305 & 9.88 &$-$0.216 & \\ 
$J16510590+1020515$ & 28.565 & 31.385 & 3.023 & 0.652 & 2.804 & 16487+1025 & 10.50 &$-$0.353 & 19.4\\ 
$J16521876-1830343$ & 1.790 & 15.928 & 3.698 & 0.740 & 3.352 & 16494$-$1825 & 3.28 &$-$0.320 & \\ 
$J16524821+0524269$ & 23.691 & 28.777 & 3.020 & 0.650 & 2.804 & 16503+0529 & 20.40 &$-$0.276 &$-$49.3\\ 
$J16534478+4857022$ & 75.428 & 39.124 & 2.938 & 0.774 & 2.543 & 16524+4901 & 16.80 &$-$0.303 & 16.7\\ 
$J17171484-0230186$ & 19.371 & 19.590 & 5.786 & 0.772 & 5.394 & 17146$-$0227 & 2.36 &$-$0.153 & \\ 
$J17210403+2655505$ & 49.629 & 30.754 & 2.450 & 0.711 & 2.146 & 17190+2658 & 33.30 &$-$0.310 & 25.8\\ 
$J17312879+3229525$ & 56.557 & 30.105 & 2.956 & 0.731 & 2.623 & 17296+3231 & 29.40 &$-$0.361 &$-$24.8\\ 
$J17331391+0820390$ & 31.591 & 21.153 & 7.726 & 2.078 & 5.454 & 17308+0822 & 12.00 & 0.021 & 8.1\\ 
$J17364445+1051070$ & 34.420 & 21.452 & 3.134 & 0.698 & 2.849 & 17343+1052 & 37.60 &$-$0.360 &$-$55.3\\ 
$J17551370+1143462$ & 37.298 & 17.715 & 6.345 & 0.818 & 5.887 & 17528+1144 & 4.80 &$-$0.184 & \\ 
$J18000391+2335371$ & 49.353 & 21.334 & 2.940 & 0.819 & 2.481 & 17579+2335 & 62.70 &$-$0.243 & 3.5\\ 
$J18024911-0632355$ & 21.477 & 7.713 & 5.468 & 0.870 & 4.935 & 18001$-$0632 & 4.51 &$-$0.157 &$-$38.8\\ 
$J18042450-0802252$ & 20.343 & 6.649 & 5.465 & 0.786 & 5.053 & 18016$-$0802 & 5.94 &$-$0.324 & \\ 
$J18080907-0649058$ & 21.870 & 6.418 & 4.941 & 0.990 & 4.235 & 18054$-$0649 & 10.60 &$-$0.118 &$-$50.4\\ 
$J18102890-0237427$ & 25.870 & 7.881 & 7.663 & 2.156 & 5.278 & 18078$-$0238 & 7.52 &$-$0.010 & 65.7\\ 
$J18105856+0753085$ & 35.444 & 12.556 & 4.220 & 0.776 & 3.823 & 18085+0752 & 21.70 &$-$0.219 &$-$64.1\\ 
$J18110144-0142340$ & 26.755 & 8.190 & 5.081 & 0.940 & 4.447 & 18084$-$0143 & 3.15 &$-$0.331 & 62.2\\ 
$J18120477-0607247$ & 22.950 & 5.888 & 4.787 & 0.859 & 4.270 & 18094$-$0608 & 6.49 &$-$0.112 & 120.6 \\ 
$J18135847-0815296$ & 21.285 & 4.464 & 5.009 & 0.793 & 4.587 & 18112$-$0816 & 6.06 &$-$0.149 & \\ 
$J18155756+0120106$ & 30.061 & 8.501 & 4.436 & 0.771 & 4.046 & 18134+0119 & 13.80 &$-$0.238 & 21.5\\ 
$J18162782-0425247$ & 24.974 & 5.723 & 5.960 & 1.499 & 4.521 & 18138$-$0426 & 20.40 & 0.013 & 58.2\\ 
$J18170265-0720564$ & 22.449 & 4.223 & 6.654 & 1.131 & 5.745 & 18143$-$0722 & 5.42 &$-$0.158 &$-$10.9\\ 
$J18182494-0331378$ & 25.999 & 5.710 & 5.745 & 0.954 & 5.091 & 18157$-$0332 & 4.91 &$-$0.273 & \\ 
$J18192465-0439593$ & 25.103 & 4.960 & 8.060 & 2.031 & 5.855 & 18167$-$0441 & 7.55 &$-$0.066 &$-$0.4\\ 
$J18193355+0354498$ & 32.797 & 8.872 & 8.763 & 2.303 & 6.167 & 18170+0353 & 5.58 &$-$0.082 & 49.6\\ 
$J18203449-0342095$ & 26.095 & 5.152 & 4.489 & 0.835 & 4.007 & 18179$-$0343 & 5.15 &$-$0.326 & 21.3\\ 
$J18205487+5031432$ & 78.759 & 25.285 & 3.224 & 0.723 & 2.903 & 18196+5030 & 27.00 &$-$0.321 &$-$0.9\\ 
$J18212420+0229022$ & 31.720 & 7.814 & 5.897 & 1.087 & 5.052 & 18188+0227 & 5.87 &$-$0.273 & 20.0\\ 
$J18213513+8238388$ & 114.678 & 27.842 & 3.029 & 0.941 & 2.394 & 18276+8236 & 44.70 &$-$0.175 &$-$26.8\\
$J18213854-0355447$ & 26.018 & 4.811 & 5.643 & 0.913 & 5.048 & 18189$-$0357 & 4.53 &$-$0.161 & 72.2\\ 
$J18223249-0305115$ & 26.871 & 5.003 & 6.132 & 1.503 & 4.688 & 18199$-$0306 & 9.43 &$-$0.022 & 55.8\\ 
$J18231986+0929569$ & 38.293 & 10.523 & 4.584 & 0.838 & 4.097 & 18209+0928 & 9.88 &$-$0.139 &$-$14.9\\ 
$J18240372+0636258$ & 35.743 & 9.080 & 8.602 & 2.545 & 5.657 & 18216+0634 & 10.10 & 0.256 & \\ 
$J18243205-0132065$ & 28.484 & 5.278 & 4.988 & 0.818 & 4.530 & 18219-0133 & 3.76 &$-$0.350 & \\ 
$J18250570-0032320$ & 29.434 & 5.610 & 4.766 & 0.755 & 4.399 & 18225$-$0034 & 7.22 &$-$0.323 & 99.1\\ 
$J18253335+0856472$ & 38.034 & 9.787 & 6.166 & 0.859 & 5.649 & 18231+0855 & 4.80 &$-$0.145 &$-$7.6\\ 
$J18264298-0024486$ & 29.736 & 5.309 & 6.508 & 1.271 & 5.398 & 18241$-$0026 & 3.66 &$-$0.142 & 124.9\\ 
$J18270226-0106095$ & 29.158 & 4.921 & 3.553 & 1.516 & 2.090 & 182444$-$0108 & 21.20 &$-$0.453 & \\ 
$J18274403+0025128$ & 30.597 & 5.465 & 6.672 & 1.654 & 5.010 & 18251+0023 & 14.40 &$-$0.105 & 99.8\\ 
$J18275079+0752206$ & 37.313 & 8.803 & 6.734 & 1.380 & 5.467 & 18254+0750 & 6.22 & 0.064 & \\ 
$J18283699+0936501$ & 38.980 & 9.404 & 6.089 & 0.820 & 5.628 & 18262+0934 & 4.50 &$-$0.123 & \\ 
$J18294430+0104534$ & 31.418 & 5.321 & 5.382 & 1.166 & 4.423 & 18271+0102 & 5.76 & 0.091 & \\ 
$J18295773+0114056$ & 31.581 & 5.341 & 7.051 & 2.191 & 4.616 & 18274+0112 & 7.31 & 0.004 & \\ 
$J18302847+0523383$ & 35.369 & 7.111 & 5.160 & 1.121 & 4.266 & 18280+0521 & 12.10 &$-$0.182 & 28.3\\ 
$J18331997+0425410$ & 34.824 & 6.041 & 4.748 & 0.852 & 4.241 & 18308+0423 & 4.88 &$-$0.329 & 72.6\\ 
$J18354242+0905384$ & 39.291 & 7.605 & 5.874 & 1.021 & 5.124 & 18333+0903 & 8.25 &$-$0.041 & 51.3\\ 
$J18364611+0845469$ & 39.110 & 7.223 & 5.306 & 1.001 & 4.585 & 18343+0843 & 4.45 &$-$0.217 &$-$9.8\\ 
$J18384242+0541298$ & 36.564 & 5.416 & 5.345 & 0.989 & 4.641 & 18362+0538 & 3.94 &$-$0.032 & 41.2\\ 
$J18414435+0802164$ & 39.010 & 5.799 & 6.551 & 0.791 & 6.132 & 18393+0759 & 3.41 &$-$0.215 & \\ 
$J18420916+0801180$ & 39.042 & 5.700 & 4.923 & 0.812 & 4.474 & 18397+0758 & 8.27 &$-$0.239 &$-$96.8\\ 
$J18421582+0731314$ & 38.609 & 5.453 & 6.968 & 1.732 & 5.194 & 18398+0728 & 8.89 & 0.055 & \\ 
$J18431907+0733137$ & 38.752 & 5.232 & 6.396 & 1.123 & 5.499 & 18408+0730 & 8.22 &$-$0.248 & \\ 
$J18454767-1148074$ & 21.774 &$-$4.118 & 4.843 & 0.929 & 4.225 & 18429$-$1151 & 6.25 &$-$0.024 & 102.5\\ 
$J18465164-1157048$ & 21.758 &$-$4.418 & 8.119 & 1.928 & 6.063 & 18440$-$1200 & 5.47 &$-$0.070 & \\ 
$J18495300-1104202$ & 22.880 &$-$4.683 & 6.355 & 0.886 & 5.799 & 18471$-$1107 & 3.09 &$-$0.164 & \\ 
$J18501116-1007570$ & 23.755 &$-$4.326 & 5.492 & 0.818 & 5.034 & 18474$-$1011 & 3.07 & 0.015 & 50.9\\ 
$J18512520+1202084$ & 43.674 & 5.460 & 5.087 & 0.869 & 4.556 & 18490+1158 & 9.06 &$-$0.024 & 35.7\\ 
$J18523817+1733113$ & 48.782 & 7.648 & 6.421 & 0.815 & 5.967 & 18504+1729 & 3.18 &$-$0.267 & 96.8\\ 
$J18524262-0951445$ & 24.278 &$-$4.759 & 4.914 & 0.753 & 4.550 & 18499$-$0955 & 5.76 &$-$0.344 & 108.0\\ 
$J18530734-1155310$ & 22.472 &$-$5.775 & 5.632 & 1.001 & 4.911 & 18503$-$1159 & 14.40 &$-$0.246 & \\  
$J18530987-1329244$ & 21.068 &$-$6.482 & 5.981 & 1.003 & 5.257 & 18503$-$1333 & 10.20 &$-$0.199 & 86.1\\ 
$J18550366-1328438$ & 21.284 &$-$6.890 & 5.543 & 0.925 & 4.931 & 18522$-$1332 & 4.94 &$-$0.117 & 28.0\\ 
$J18554765-1415207$ & 20.661 &$-$7.393 & 6.154 & 1.170 & 5.189 & 18529$-$1419 & 3.80 &$-$0.006 & 83.3\\ 
$J18562524-0744227$ & 26.594 &$-$4.622 & 5.425 & 1.350 & 4.201 & 18537$-$0748 & 7.80 &$-$0.026 & \\ 
$J18572211-0831208$ & 25.998 &$-$5.184 & 6.518 & 1.321 & 5.336 & 18546$-$0835 & 6.15 &$-$0.004 & 47.6 \\ 
$J18572648+1349096$ & 45.936 & 4.950 & 8.039 & 2.776 & 4.762 & 18551+1345 & 57.80 &$-$0.261 & 76.2 \\ 
$J18574985+2031371$ & 52.022 & 7.854 & 5.634 & 1.203 & 4.622 & 18556+2027 & 14.60 &$-$0.358 & \\ 
$J18575917+1413196$ & 46.357 & 5.013 & 7.449 & 1.777 & 5.610 & 18556+1409 & 5.72 &$-$0.090 & 86.4\\ 
$J18594368-0924127$ & 25.469 &$-$6.099 & 5.539 & 0.841 & 5.048 & 18569$-$0928 & 7.19 &$-$0.233 & 40.6\\ 
$J19004752-0742491$ & 27.106 &$-$5.577 & 5.072 & 0.993 & 4.362 & 18580$-$0747 & 23.20 &$-$0.140 & 84.7\\ 
$J19010471-1050004$ & 24.330 &$-$7.034 & 7.288 & 1.943 & 5.210 & 18582$-$1054 & 13.70 &$-$0.047 & 28.2\\ 
$J19010944+1538566$ & 47.982 & 4.972 & 5.092 & 0.786 & 4.680 & 18588+1534 & 8.74 &$-$0.372 & 24.2\\ 
$J19012574-0529398$ & 29.167 &$-$4.719 & 4.629 & 0.837 & 4.144 & 18587$-$0534 & 13.90 &$-$0.214 & 87.3\\ 
$J19020569-1236483$ & 22.830 &$-$8.045 & 4.918 & 0.810 & 4.472 & 18593$-$1241 & 6.48 &$-$0.145 &$-$21.0\\ 
$J19024407-0611124$ & 28.694 &$-$5.321 & 5.444 & 0.761 & 5.068 & 19000$-$0615 & 6.18 &$-$0.036 & 95.4\\ 
$J19030151+1656312$ & 49.343 & 5.155 & 6.797 & 0.949 & 6.150 & 19007+1652 & 4.30 &$-$0.072 & \\ 
$J19034074-1409330$ & 21.596 &$-$9.068 & 6.465 & 1.130 & 5.558 & 19008$-$1414 & 6.15 &$-$0.162 & \\ 
$J19050061+2310298$ & 55.161 & 7.529 & 4.800 & 0.778 & 4.400 & 19029+2305 & 6.41 &$-$0.167 &$-$22.0\\ 
$J19054911-1212243$ & 23.604 &$-$8.682 & 6.395 & 1.407 & 5.089 & 19030$-$1217 & 9.94 &$-$0.132 &$-$16.6\\ 
$J19060940+1738007$ & 50.302 & 4.801 & 7.520 & 2.093 & 5.226 & 19039+1733 & 3.92 &$-$0.165 & \\ 
$J19062416+1739313$ & 50.351 & 4.760 & 5.891 & 0.858 & 5.375 & 19041+1734 & 6.30 &$-$0.185 &   \\ 
$J19062439-1509534$ & 20.967 &$-$10.099 & 6.576 & 1.254 & 5.490 & 19035$-$1514 & 3.82 &$-$0.031 &$-$48.3\\ 
$J19074038-0515161$ & 30.087 &$-$5.998 & 6.476 & 1.242 & 5.408 & 19050$-$0520 & 3.14 &$-$0.045 & 99.0\\ 
$J19085920-1510032$ & 21.237 &$-$10.662 & 5.867 & 1.054 & 5.069 & 19061$-$1514 & 8.31 &$-$0.108 & 46.7\\ 
$J19090454+2939292$ & 61.454 & 9.565 & 2.130 & 0.622 & 1.954 & 19071+2934 & 26.10 &$-$0.207 &$-$13.1\\ 
$J19094939-0804034$ & 27.795 &$-$7.736 & 5.058 & 0.908 & 4.470 & 19071$-$0808 & 6.79 &$-$0.132 & 30.4\\ 
$J19102783-1541355$ & 20.909 &$-$11.209 & 5.746 & 0.777 & 5.347 & 19075$-$1546 & 3.79 &$-$0.348 & \\ 
$J19105141-0328410$ & 32.041 &$-$5.903 & 6.061 & 0.905 & 5.478 & 19082$-$0333 & 5.54 &$-$0.309 & \\ 
$J19111079-0227493$ & 32.986 &$-$5.514 & 6.493 & 1.251 & 5.412 & 19085$-$0232 & 3.46 &$-$0.041 & 51.7\\ 
$J19134664-0326081$ & 32.411 &$-$6.533 & 5.068 & 0.833 & 4.588 & 19111$-$0331 & 8.63 &$-$0.131 & \\ 
$J19141725-0850549$ & 27.583 &$-$9.072 & 4.819 & 0.803 & 4.383 & 19115$-$0856 & 6.25 &$-$0.296 & \\ 
$J19145484-0225287$ & 33.447 &$-$6.328 & 8.125 & 1.988 & 5.982 & 19122$-$0230 & 9.29 & 0.304 & \\ 
$J19173939-1322488$ & 23.802 &$-$11.792 & 4.712 & 0.821 & 4.250 & 19148$-$1328 & 9.22 &$-$0.201 & \\ 
$J19181178-0721437$ & 29.367 &$-$9.279 & 6.395 & 0.831 & 5.918 & 19155$-$0727 & 3.14 &$-$0.225 & \\ 
$J19182271-0242108$ & 33.593 &$-$7.225 & 7.020 & 1.185 & 6.034 & 19157$-$0247 & 8.88 &$-$0.093 & \\ 
$J19211169+0320578$ & 39.326 &$-$5.071 & 6.519 & 1.626 & 4.898 & 19186+0315 & 15.50 & 0.088 &$-$22.2\\ 
$J19222258-1418050$ & 23.449 &$-$13.221 & 5.598 & 1.400 & 4.302 & 19195$-$1423 & 8.78 & 0.170 & \\ 
$J19233466+0037583$ & 37.181 &$-$6.855 & 5.395 & 0.778 & 4.995 & 19210+0032 & 5.96 &$-$0.350 & 26.2\\ 
$J19234517+7141137$ & 103.152 & 23.178 & 2.836 & 0.989 & 2.132 & 19243+7135 & 48.50 &$-$0.349 & 14.0\\ 
$J19240522-0722442$ & 30.010 &$-$10.595 & 4.897 & 0.776 & 4.500 & 19213$-$0728 & 8.75 &$-$0.253 & 31.2\\ 
$J19265373-0104251$ & 36.036 &$-$8.376 & 5.682 & 1.000 & 4.962 & 19243$-$0110 & 9.73 &$-$0.166 & \\ 
$J19340281+0926061$ & 46.225 &$-$5.022 & 6.822 & 1.347 & 5.602 & 19316+0919 & 6.06 &$-$0.263 & 92.1\\ 
$J19341153+1958285$ & 55.483 & 0.036 & 4.004 & 0.771 & 3.614 & 19320+1951 & 27.40 &$-$0.265 & 45.8\\ 
$J19360670+0945041$ & 46.748 &$-$5.317 & 5.789 & 0.877 & 5.246 & 19337+0938 & 3.34 &$-$0.199 & 17.4\\ 
$J19410829+0202312$ & 40.505 &$-$10.086 & 6.011 & 1.058 & 5.207 & 19386+0155 & 17.40 & 0.435 & \\ 
$J19433664+1049180$ & 48.588 &$-$6.412 & 7.929 & 1.570 & 6.388 & 19412+1042 & 3.89 &$-$0.030 & \\ 
$J19433804+1403158$ & 51.421 &$-$4.830 & 5.367 & 0.788 & 4.952 & 19413+1356 & 3.89 &$-$0.262 & 28.4\\ 
$J19444164+0513039$ & 43.769 &$-$9.360 & 4.806 & 0.751 & 4.445 & 19422+0505 & 5.33 &$-$0.386 & 35.6\\ 
$J19464724+1549014$ & 53.336 &$-$4.618 & 7.676 & 1.549 & 6.165 & 19445+1541 & 3.76 &$-$0.098 & 56.8\\ 
$J19483842+2759358$ & 64.082 & 1.141 & 3.685 & 0.607 & 3.531 & 19466+2751 & 27.80 &$-$0.011 & \\ 
$J19492572+0231306$ & 41.935 &$-$11.683 & 7.208 & 0.790 & 6.790 & 19469+0223 & 4.79 &$-$0.061 & \\ 
$J19510953+1354375$ & 52.207 &$-$6.486 & 5.909 & 0.758 & 5.537 & 19488+1346 & 5.38 &$-$0.154 & 40.2\\ 
$J19542885+1943430$ & 57.650 &$-$4.222 & 5.547 & 1.555 & 4.028 & 19522+1935 & 29.30 &$-$0.055 & 66.7\\ 
$J19552508+0156036$ & 42.133 &$-$13.280 & 5.741 & 1.338 & 4.534 & 19528+0148 & 51.50 &$-$0.164 & \\ 
$J19553800-0018428$ & 40.121 &$-$14.390 & 5.971 & 0.768 & 5.585 & 19530-0026 & 5.18 &$-$0.313 & \\ 
$J19554049+1805361$ & 56.385 &$-$5.302 & 6.059 & 1.312 & 4.890 & 19534+1757 & 5.29 & 0.101 & \\ 
$J19595639+1151450$ & 51.510 &$-$9.366 & 5.159 & 0.774 & 4.764 & 19575+1143 & 10.20 &$-$0.241 & \\ 
$J20003856+1331331$ & 53.048 &$-$8.667 & 5.143 & 0.792 & 4.723 & 19583+1323 & 14.20 &$-$0.192 &$-$25.4\\ 
$J20030250+0544166$ & 46.496 &$-$13.097 & 5.855 & 0.792 & 5.435 & 20005+0535 & 7.16 &$-$0.112 & \\ 
$J20042163+1748345$ & 57.217 &$-$7.216 & 4.771 & 0.821 & 4.309 & 20020+1739 & 8.51 &$-$0.215 & 27.7\\ 
$J20075461+1842544$ & 58.441 &$-$7.456 & 7.471 & 1.493 & 6.041 & 20056+1834 & 17.50 & 0.012 & \\ 
$J20080544+1516428$ & 55.511 &$-$9.303 & 7.593 & 1.326 & 6.404 & 20057+1507 & 3.28 &$-$0.194 & \\ 
$J20120916+1116516$ & 52.564 &$-$12.232 & 6.824 & 1.793 & 4.962 & 20097+1107 & 6.47 & 0.005 &$-$13.9\\ 
$J20172312+1718459$ & 58.463 &$-$10.109 & 5.710 & 0.907 & 5.124 & 20151+1709 & 8.03 &$-$0.237 & 5.1\\ 
$J20292219+1311116$ & 56.526 &$-$14.779 & 4.939 & 0.900 & 4.363 & 20270+1301 & 7.86 &$-$0.270 & 24.2\\ 
$J20571628+0258445$ & 51.363 &$-$26.113 & 8.405 & 1.727 & 6.638 & 20547+0247 & 45.50 &$-$0.129 & \\ 
\end{longtable}

%
\tabcolsep 3pt
\begin{longtable}{llrrrrrrrr}
\caption{Candidates for the deviant motions ($|b|>3^{\circ}$).}
\hline
2MASS    & IRAS &   $l$       &    $b$      &  $v_{\rm LSR}$  & $K$  & $H-K$ & $K_c$ & $F_{12}$ & $C_{12}$ \\
name &     name      & ($\circ$)  &  ($\circ$)  &   (km s$^{-1}$) &      &       &       &  (Jy)    &          \\
\hline
\endfirsthead
\hline
2MASS    & IRAS &   $l$       &    $b$      &  $v_{\rm LSR}$  & $K$  & $H-K$ & $K_c$ & $F_{12}$ & $C_{12}$ \\
name &     name      & ($\circ$)  &  ($\circ$)  &   (km s$^{-1}$) &      &       &       &  (Jy)    &          \\
\hline
\endhead
\hline
\endfoot
\hline
\multicolumn{10}{l}{$^{*}$ detection in this work.} \\
\multicolumn{10}{l}{$^{\dagger}$ Hipparcos proper motion data exists.} \\
\endlastfoot
$J16254746+1853328^{\dagger}$ & 16235$+$1900  &   35.35 &   40.35  &  $-15.0$ &$-0.63$&  1.09 &$-0.49$& 499.8 &  $-0.44$ \\ 
$J16524821+0524269^*$ & 16503$+$0529  &   23.69 &   28.78  &  $-49.3$ &  3.02 &  0.65 &  2.80 &  20.4 &  $-0.28$ \\
$J17071761+1710228$ & 17050$+$1714  &   37.70 &   30.52  &  $-76.6$ &  3.06 &  0.54 &  3.00 &   7.6 &  $-0.22$ \\
$J17081033-0220225$ & 17055$-$0216  &   18.30 &   21.63  &  $-40.6$ &  6.21 &  1.11 &  5.34 &   6.4 &  $-0.08$ \\
$J17354000+1535122$ & 17334$+$1537  &   38.98 &   23.63  &  $-53.0$ &  1.69 &  0.88 &  1.14 & 154.0 &  $-0.18$ \\
$J17364445+1051070^*$ & 17343$+$1052  &   34.42 &   21.45  &  $-55.3$ &  3.13 &  0.70 &  2.85 &  37.6 &  $-0.36$ \\
$J17450262-0512365$ & 17423$-$0511  &   20.49 &   12.23  &  $-30.6$ &  3.62 &  0.58 &  3.50 &   8.8 &  $-0.30$ \\
$J18024911-0632355^*$ & 18001$-$0632  &   21.48 &    7.71  &  $-38.8$ &  5.47 &  0.87 &  4.94 &   4.5 &  $-0.18$ \\
$J18080907-0649058^*$ & 18054$-$0649  &   21.87 &    6.42  &  $-50.4$ &  4.94 &  0.99 &  4.24 &  10.6 &  $-0.12$ \\
$J18105856+0753085^*$ & 18085$+$0752  &   35.44 &   12.56  &  $-64.1$ &  4.22 &  0.78 &  3.82 &  21.7 &  $-0.22$ \\
$J18163694+0341352^{\dagger}$ & 18141$+$0340  &   32.26 &    9.43  &  $-49.0$ &  2.70 &  0.49 &  2.72 &   7.8 &  $-0.43$ \\
$J18364919-1351561$ & 18339$-$1354  &   18.94 & $-3.11$  &  $-36.9$ &  3.60 &  0.59 &  3.47 &  13.8 &  $-0.37$ \\
$J18382111+0850030^{\dagger}$ & 18359$+$0847  &   39.35 &    6.90  &  $-57.0$ &$-0.82$&  0.46 &$-0.76$& 409.0 &  $-0.45$ \\
$J18420916+0801180^*$ & 18397$+$0758  &   39.04 &    5.70  &  $-96.8$ &  4.92 &  0.81 &  4.47 &   8.3 &  $-0.24$ \\ 
$J18475335-0919102$ & 18450$-$0922  &   24.23 &  $-3.45$ &  $-49.4$ &  4.03 &  1.41 &  2.71 &  80.3 &  $-0.14$ \\
$J19062439-1509534^*$ & 19035$-$1514  &   20.97 & $-10.10$ &  $-48.3$ &  6.58 &  1.25 &  5.49 &   3.8 &  $-0.03$ \\
$J19222258-1418050^*$ & 19195$-$1423  &   23.45 & $-13.22$ &  $-74.6$ &  5.60 &  1.40 &  4.30 &   8.8 &  $ 0.17$ \\
$J19291840-1916199$ & 19263$-$1922  &   19.47 & $-16.78$ &  $-69.3$ &  4.22 &  0.57 &  4.12 &  11.9 &  $-0.19$ \\
$J19310440-1640135$ & 19281$-$1646  &   22.12 & $-16.11$ &  $-69.5$ &  5.05 &  1.09 &  4.21 &  24.7 &  $-0.11$ \\
$J20261323-1347584$ & 20234$-$1357  &   30.72 & $-27.14$ &  $-33.3$ &  3.24 &  0.98 &  2.55 &  46.7 &  $-0.02$ \\
\end{longtable}
\clearpage
\begin{table}[htbp]
  \caption{Parameters used for the bar model}\label{tab:table6}
    \begin{center}
  \begin{tabular}{lr}
  \hline\hline
parameter  & value \\   
	\hline
rotational velocity of the LSR  & 220 km s$^{-1}$\\
Galactocentric distance of the Sun & 8 kpc \\ 
pattern speed of the bar & 55 km s$^{-1}$ kpc$^{-1}$\\
angle between bar major axis and the Sun-GC line & 30$^{\circ}$\\
radius of the corotation  & 4 kpc\\
radius of the outer Lindblad resonance & 6.8 kpc\\
strength of the bar ($\epsilon/2$) & 0.025 \\
bar damping constant & 4.1 km s$^{-1}$ kpc$^{-1}$\\
\hline
\end{tabular}
\end{center}
\end{table}
\begin{table}[htbp]
  \caption{Observational results of H$_2$O Masers.} 
      \begin{center}
  \begin{tabular}{lrrrrr}
\hline  \hline
   & \multicolumn{4}{}{$J=1$--0 $v=1$}  &  \\
2MASS name &  $T_a$  &   $V_{\rm lsr}$  & L.F.  &  $rms$   & obs. date \\  
   & {\tiny (K)} & {\tiny (km s$^{-1}$)} & {\tiny (K km s$^{-1}$)} &  {\tiny (K)}  & {\tiny (yymmdd.d)}\\
\hline
$J12370691-1731319$ &  ...   &   ...   & ...   & 0.066 & 090530 \\
$J12583891+2308215$ &  1.967 &   26.5  & 3.077 & 0.117 & 090530 \\
$J15284369+0349430$ &  1.683 &   44.7  & 1.472 & 0.102 & 090530 \\
$J16521876-1830343$ &  ...   &  ...    & ...   & 0.094 & 090417 \\
$J17331391+0820390$ &  ...   &  ...    & ...   & 0.095 & 090417 \\
$J18080907-0649050$ &  0.345 & $-56.5$ & 1.333 & 0.063 & 090417 \\
$J18155756+0120106$ &  ...   &  ...    & ...   & 0.065 & 090417 \\
\hline
\end{tabular}
\end{center}
\end{table}
%

\onecolumn

\clearpage


\end{document}